\documentclass[onecolumn,aps,longbibliography]{revtex4-2}

\usepackage{fleqn,graphicx}
\usepackage[normalem]{ulem}
\usepackage[dvipsnames]{xcolor}
\usepackage{color,soul}
\soulregister\cite7
\soulregister\ref7

\setstcolor{Green}
\usepackage[caption=false]{subfig}

\usepackage{amsmath,amsfonts,amssymb}
\usepackage{braket}
\begin{document}

\title {Effects of Markovian Noise and Cavity Disorders on the Entanglement Dynamics of Double Jaynes-Cummings Models}

\author{Harsh Rathee}
\affiliation{QOLS, Blackett Laboratory, Imperial College London, London SW7 2BW, United Kingdom}
\author{Kishore Thapliyal}
\affiliation{Joint Laboratory of Optics, Faculty of Science,
Palack\'{y} University, Czech Republic, 17. listopadu 12, 779~00
Olomouc, Czech Republic}
\author{Anirban Pathak}
\email{anirban.pathak@jiit.ac.in}
\affiliation{Department of Physics and Materials Science \& Engineering, Jaypee Institute of Information Technology, A 10, Sector 62, Noida, UP 201309, India}
\keywords{Double Jaynes-Cummings models, Entanglement dynamics, Entanglement sudden death, Markovian noise, Noisy Jaynes-Cummings model}

\begin{abstract}

The ability to prepare and manipulate non-classical states, such as entangled qubits, is fundamental to the development of quantum information processing, communication, and computation. In this work, we investigate the dynamics of a double Jaynes-Cummings model, a well-established theoretical framework for studying light-matter interactions that captures essential features of a wide range of quantum systems, including circuit QED, optomechanics, and atomic cavity systems. We examine the model under the influence of Markovian noise and static (glassy) cavity disorder. The study aims to elucidate the impact of these imperfections on entanglement dynamics. The system is initialized with the cavity fields in vacuum and the two atoms in a specific entangled superposition state. Through numerical simulations, we observe that the presence of noise and nonlinear pumping gives rise to nontrivial features in the entanglement evolution, including the emergence of entanglement sudden death (ESD) and subsequent revivals in scenarios where such phenomena are absent in the idealized model. Markovian noise leads to a monotonic decay of entanglement, while disorder tends to wash out the entanglement features. Nonlinear interactions, on the other hand, accelerate the dynamical evolution. The combined and competing effects of noise, disorder, and nonlinearity are systematically analyzed, revealing rich and intricate behavior in the entanglement dynamics. These results contribute to a deeper understanding of the robustness and control of entanglement in open quantum systems with imperfections, which is essential for realistic implementations of quantum technologies.

% The dynamics of double Jaynes-Cummings models are numerically studied in the presence of Markovian noise and cavity disorders with specific attention to entanglement sudden death and revivals. The study is focused on the glassy disorders, which remain unchanged during the observations. The field is initially assumed to be in a vacuum state, while the atoms are considered to be in a particular two-qubit superposition state. Specifically, the study has revealed that the presence of noise, or a nonlinear pump results in interesting behaviors in the entanglement dynamics. Further, entanglement sudden death is observed in the presence of Markovian noise and a nonlinear pump. The presence of entanglement sudden deaths and revivals have also been observed in cases where they were absent initially for the chosen states. The effect of noise on the dynamics of the system is to decay the characteristics, while that of the disorder is to wash them out. On the other hand, the introduction of nonlinearity causes a speed-up of the dynamics of the system. The combined effects of these dynamics are also studied. The competing effects of disorder and noise in washing out the entanglement characteristics are compared, and the effect of nonlinearity on the competition dynamics is studied.

\end{abstract}

\maketitle
\section {Introduction}

Interest in the field of quantum computation and quantum information theory has been rising rapidly in recent years. The advantage that the power of quantum mechanics brings to the fields of computation, communication, and metrology has been well established by the works of Feynman~\cite{feynman1986quantum, feynman1982simulating}, Bennett and Brassard~\cite{BB84}, Deutsch~\cite{deutsch1985quantum}, and Shor~\cite{shor1994algorithms} among others. An increasing interest is emerging in solidifying this phenomenon of \textit{quantum advantage} and the study of its utilization in a diverse range of fields~\cite{arute2019quantum,nagali2012experimental, kumar2019experimental, centrone2021experimental, bravyi2020quantum, novo2021quantum, maslov2021quantum}. 
% While a lot of the proposed applications rely on the lofty goal of building a fault-tolerant quantum computer, there have been demonstrated use cases for noisy intermediate-scale quantum (NISQ) devices as well. These range from simulating many-body physics~\cite{SKP+2019} and studying open quantum systems~\cite{MKM+2021} to the study of fundamental symmetries in physics~\cite{DMP+2021}. 

There are various criteria that the underlying architecture for a quantum computer must be able to follow to allow for universal quantum computation~\cite{QCQI}. One of these criteria is the ability to prepare fiducial initial states, especially nonclassical states (i.e., states that do not have a classical analog). Proper utilization of these nonclassical states is one of the ways quantum computing can provide this advantage over its classical counterpart~\cite{H2013, W2013}. Of these states, one of particular interest is the class of entangled states. One common way of generating these entangled states is through a rather simple process of interaction of a two-level atom with a monochromatic light field described by the Jaynes-Cummings (JC) model. This model has been studied extensively~\cite{FLR+2019,FLZ2020,HLK2020,CQW+2021,WZ2020,akella2022dynamics,greentree2013special} since its development in 1963~\cite{jaynes1963comparison,P1963}.

Interestingly, the JC model is still relevant and an important tool for studying many aspects of quantum information processing. It forms the cornerstone of cavity quantum electrodynamics (QED), which is used to describe the interaction of qubits with driving microwave fields in the study of superconducting qubits~\cite{CBH+2004, AIN+2007}. It has also been widely used for the generation of engineered quantum states~\cite{dell2006multiphoton} and the study of non-Markovian evolution of open quantum systems~\cite{bellomo2007non}. The JC model (in its original form, as well as various generalized forms~\cite{ermann2020jaynes, GDS+20, BR19, HLK2020}) is a quite effective way to test and study various quantum phenomena including, but not limited to, generation of Schr\"odinger-cat states~\cite{CQW+2021, buzek1992schrodinger}, entanglement protection~\cite{fasihi2019entanglement}, catalysis~\cite{Messinger_2020}, multiphoton blockade~\cite{zou2020multiphoton}, fractional revivals~\cite{averbukh1992fractional}, quantum state engineering~\cite{gea1990collapse}, strong squeezing~\cite{kuklinski1988squeezing}, entanglement generation~\cite{phoenix1991entangled}, state discrimination~\cite{namkung2019almost}, coherent control~\cite{LLM2024}, quantum phase transitions~\cite{HYM+2024}. Applications of the JC model are not restricted to the above-listed examples, and interested readers may review the recent article by Jaoas et al.~\cite{JTS+2024} (and references therein) to appreciate the relevance of the JC model.

Motivated by the above facts, here we plan to study the rich dynamics of entangled atoms in cavities under the influence of noise due to interaction with the ambient environment. This investigation will help us to understand how well the resource entangled states retain their entanglement properties under noisy conditions, and determine whether working in certain noise regimes allows the entanglement resource to be revived. We also investigate whether introducing a nonlinear atom-field interaction~\cite{di2012extremely} enhances the system’s robustness against noise. Additionally, we delve into the role of nonlinearity in the driving field in introducing entanglement deaths and revivals in the dynamics of the system. Along with dissipations, disorders are also unavoidable in real-life systems. Their independent effects on the dynamics have been studied in previous works~\cite{GDS+20, BR19}. However, the combined effect of these deteriorating factors is not known. In this paper, we aim to study the combined effects of these factors on entanglement dynamics. Since these mechanisms are present in almost all practical situations, a study of their combined effects on entanglement dynamics can guide the efficient use of quantum systems. The results can also be utilized to guide the efficient engineering of physical systems that are robust to noise and disorder. 

We focus on the Double JC model, which has been a topic of interest in various previous studies~\cite{GDS+20, YE2004, YYE06, CRF2009, VJF+2011, K2011, SMA+2008, SB2007, HFC+2008, CYL+2010, OCF2011, ZX2010, SSW+2017, PDR+2018}. In Section~\ref{sec:JCmodel}, we discuss linear and nonlinear double JC models with significant attention to nonlinear JC models involving multiphoton interaction and nonlinear driving field. Inclusion of nonlinear interaction terms in the study is inspired by the nonlinear Hamiltonians that appear in physical systems like trapped ions~\cite{WR95, DCB+96}, optical~\cite{DQS+92} and microwave cavities~\cite{MJP+87}, optomechanical systems~\cite{JMA+08, vanner11}, and even superconducting circuits~\cite{LRV+15, SDE+18, SMA18}. Origin and possible ways of modeling various types of disorder and noise have been discussed in Section~\ref{sec:disorder&noise}. Section~\ref{sec:results} discusses the results and analysis of the dynamics of our system with special attention given to the impact of noise and disorder on entanglement (produced via the JC type interaction), leading to its sudden death, followed by sudden revivals.

\section{System and mathematical description}\label{sec:math}

\subsection{The Jaynes-Cummings Model and its generalizations}\label{sec:JCmodel}

The interaction of a two-level quantum system (like an atom) with a quantized electromagnetic field is described by the JC Hamiltonian. In the simplest form of the JC model, an atom (or equivalently a two-level quantum system) interacts with the field through a dipole interaction term given by $- {\hat{d}\cdot\hat{E}}$, where ${\hat d}$ is the dipole moment operator and ${\hat{E}}$ is the electric field operator. The total Hamiltonian can be defined as~\cite{shore1993jaynes, gerry2005introductory}

\begin{equation}
    \hat{\mathcal{H}} = \frac{1}{2} \hbar \omega_0
    \hat{\sigma}_3 + \hbar \omega \hat{a}^\dagger \hat{a} + \hbar g (\hat{\sigma}_+ + \hat{\sigma}_{-})(\hat{a}+\hat{a}^\dagger),
    \label{eq:singleJC}
\end{equation}

Here, $\hat{\sigma}_+ = \ket{e}\bra{l}$ and $\hat{\sigma}_{-} = \ket{l}\bra{e}$ with $\ket{l}$ and $\ket{e}$ representing the ground and the excited states of the atom, respectively. $\hat{\sigma}_3 = \ket{e}\bra{e} - \ket{l}\bra{l}$, and $\hat{a}$ is the annihilation operator of the optical field. The frequency of the optical (electric) field is represented by $\omega$, and $\omega_0$ is the frequency of the transition between the ground and excited states of the atom. Also, $g$ is the coupling strength that parameterizes the interaction of the atom with the field in the cavity. Using the rotating wave approximation (RWA), the Hamiltonian in the interaction picture (rotating frame of the field) is given by,~\cite{gerry2005introductory}

\begin{equation}
    \hat{\mathcal{H}}_{I,RWA} = (\omega_0 - \omega) \frac{1}{2} \hbar
    \hat{\sigma}_3 + \hbar g (\hat{\sigma}_+ \hat{a} + \hat{\sigma}_{-} \hat{a}^{\dagger}).
    \label{eq:singleJC,I,RWA}
\end{equation}

The dynamics of the JC model given by Eq.~(\ref{eq:singleJC,I,RWA}) can be solved analytically~\cite{gerry2005introductory}. It is also possible to find approximate solutions to more generalized versions of the model, like one coupled to a driven optomechanical system~\cite{LIJ2020}.  

This simple model of matter-field interaction can be generalized in various ways. There are many variants of the JC model that are obtained under various assumptions (see~\cite{DAS2024}). One such possibility is to consider that two atoms are interacting with two uncoupled cavities. Such a consideration leads to the double JC model and its generalizations to be discussed here.

\subsubsection{Double JC model}

The Hamiltonian of the double JC model under RWA, in the case of non-interacting cavities, can be obtained easily using Eq.~(\ref{eq:singleJC}). It is simply a direct sum of two JC Hamiltonians and the same may be explicitly written as~\cite{GDS+20}

\begin{equation}\label{eq:doubleJC,RWA}
\hat{\mathcal{H}}_{RWA} = \frac{1}{2} \hbar \omega_0 \hat{\sigma}_3^A + \hbar \omega \hat{a}^\dagger \hat{a} + \hbar G_A (\hat{\sigma}_+^A \hat{a} + \hat{\sigma}_-^A \hat{a}^{\dagger}) + \frac{1}{2} \hbar \omega_0 \hat{\sigma}_3^B + \hbar \omega \hat{b}^\dagger \hat{b} + \hbar G_B (\hat{\sigma}_+^B \hat{b} + \hat{\sigma}_-^B \hat{b}^{\dagger}). 
\end{equation}

Here, $\omega_0$ is the frequency associated with the $\ket{l} \rightarrow \ket{e}$ transition for both atoms. $\hat{a}\,(\hat{a}^{\dagger})$ and $\hat{b}\,(\hat{b}^{\dagger})$ are the annihilation (creation) operators for the single mode fields coupled to atoms $A$ and $B$ with coupling constants $G_A$ and $G_B$, respectively. The superscripts $A$ and $B$ on the raising and lowering operators refer to the two atoms. The double JC model is relevant for studying the evolution of quantum properties between the two involved atoms~\cite{YE2004, YYE06, CRF2009, VJF+2011, K2011}. One of the ways to extend the discussion of the double JC model to study physical systems is to include interaction between the cavities. Under certain assumptions, it is possible to approximately solve the dynamics of that system \cite{IAJ+2020}.

\subsubsection{Multi-photon double JC models}

We can extend the single and double JC models to study multi-photon interactions and refer to such JC models as multi-photon JC models. Introduction of $N$-photon interaction in the double JC model transforms the Hamiltonian in Eq.~(\ref{eq:doubleJC,RWA}) to~\cite{shore1993jaynes, BR19}

\begin{equation}
    \hat{\mathcal{H}}_{RWA} = \frac{1}{2} \hbar \omega_0
    \hat{\sigma}_3^A + \hbar \omega \hat{a}^\dagger \hat{a} + \hbar G_A (\hat{\sigma}_+^A \hat{a}^N + \hat{\sigma}_-^A (\hat{a}^N)^{\dagger}) + \frac{1}{2} \hbar \omega_0
    \hat{\sigma}_3^B + \hbar \omega \hat{b}^\dagger \hat{b} + \hbar G_B (\hat{\sigma}_+^B \hat{b}^N + \hat{\sigma}_-^B (\hat{b}^N)^{\dagger}).
    \label{eq:doubleJC,non-lin,RWA}
\end{equation}

Here, the interaction may be visualized as a physical process in which the transition of an atom from the ground to an excited (excited to ground) state happens in $N$ steps by absorbing (emitting) $N$ photons of equal frequency.  Realization of such multi-photon optical processes became possible with the advent of the laser, and now these processes are frequently realized in laboratories (see~\cite{BR19} for details). For the resonant case, we must have $\omega_0 = N\omega$. Another possible route for generalization or extension of the JC model arises from considering a nonlinearly interacting driving field. 

\subsubsection {JC model in the presence of a nonlinearly interacting driving field}

The cavity may be driven externally by an interaction of a probe field with either the cavity field or atom. In the case of high-intensity lasers, this probe field may also interact with the cavity field through a nonlinear process~\cite{di2012extremely}. The nonlinear driving field can be described by  $\hat{\mathcal{H}}_P = \epsilon(t)[\hat{a}^Me^{i(M\omega_Pt-\chi)} + \mathrm{H.c.}]$. We use ${\epsilon(t)}$ to refer to the strength of the driving field. Here, $\hat{a}$ is the cavity mode, $\chi$ is the phase, $M$ represents the nonlinearity of the field, and $\omega_P$ is the single photon frequency. $M=1$ yields the linearly driven cavity case. 

Inclusion of the pump changes the Hamiltonian of the double JC model in the interaction picture (rotating frame of the pump) to~\cite{BR19}

\begin{equation}
    \hat{\mathcal{H}}_{I,RWA} = \Delta_P(\hat{a}^\dag\hat{a} + \hat{b}^\dag\hat{b} +  N\frac{\sigma_3^A}{2} + N\frac{\sigma_3^B}{2})
    + [\epsilon(t)e^{-i\chi}(\hat{a}^M + \hat{b}^M) + G_A\hat{a}^N\sigma^A_+ + G_B\hat{b}^N\sigma^B_+ +{\mathrm{H.c.}}],
    \label{eq:doubleJC,non_lin,pump}
\end{equation} 

where $\Delta_P = \omega-\omega_P$ and ${\mathrm{H.c.}}$ corresponds to the Hermitian conjugate terms. The resonant case is $\omega_P = \omega$ when $M<N$, and $N\omega_P = N\omega \pm G_{\{A,B\}}\sqrt{N!}$ when $M = N$.

So far, we have considered some ideal descriptions of double JC models. We now discuss a way of quantifying the atom-atom entanglement as we are interested in the dynamics of this entanglement in the present study. In what follows, we include some imperfections in the model, such as cavity disorders or noise. 

\subsection{Atom-atom entanglement and initial atom-cavity states}

For an arbitrary two-qubit mixed state, concurrence captures the entanglement as a monotonically increasing function~\cite{HW97}. Hence, we use concurrence as a measure of the qubit-qubit entanglement. Concurrence is calculated using the formula 

\begin{equation}
C = \max\left\{0,\sqrt{\lambda_1} - \sqrt{\lambda_2} - \sqrt{\lambda_3} - \sqrt{\lambda_4}\right\}, \label{eq:conC}
\end{equation}

where, $\lambda_1,\, \lambda_2,\, \lambda_3,$  and $\lambda_4$ are the eigenvalues of matrix $\rho_{AB}\tilde{\rho}_{AB}$ (arranged in the descending order) with 

\begin{equation}
\tilde{\rho}_{AB} = (\sigma_y \otimes \sigma_y)\rho^*_{AB}(\sigma_y \otimes \sigma_y)
\end{equation}

$\sigma_y$ being the Pauli operator. $\rho_{AB}$ is the two-qubit mixed state given by ${\mathrm {Tr}}_{\mathrm{cav}} [\rho_{\mathrm{sys}}(t)]$.

Once the entanglement is quantified through a measure (concurrence in the present case), its dynamics can be studied to observe entanglement abruptly becoming zero for finite times. This vanishing of entanglement at a finite time, for certain parameters, with a discontinuous derivative is referred to as the entanglement sudden death. This phenomenon is sometimes followed by the revival of entanglement in the system with a discontinuous gradient, known as sudden revival. An experimental proposal~\cite{MPL+2006} for observing entanglement sudden death was soon followed by experimental realization in an optical system~\cite{MFM+2007} and atomic ensembles~\cite{JKH+2007}. It has also been predicted to arise in hybrid quantum systems of a fiber atom optomechanical system~\cite{MHM+2022}. Even then, this phenomenon of entanglement of a system subjected to a reservoir dying out in a finite time while the local coherences decay asymptotically is considered a surprising effect. Analytical models for a two-qubit system interacting with a reservoir showing the entanglement sudden death have been studied by Yu and  Eberly~\cite{YE2004, TJ2006, JT2007}, while a geometric analysis was performed by Cunha~\cite{cunha2007}. Entanglement is an important resource for many quantum protocols, such as quantum teleportation, quantum computation, and quantum key distribution~\cite{QCQI, CGC+1993, horodecki2009quantum, NDS+2014}. The existence of noise causes the quantum coherences to decay at an exponential rate, but as eluded to before, in certain cases, entanglement can vanish in a finite time. Thus, it is of interest to find ways and protocols to avoid this phenomenon and to enhance the frequency and possibility of entanglement revivals~\cite{NS2019}. Dynamics of entanglement are already studied for different variants of JC models, including a double JC model with squeezed and thermal photons~\cite{KCM2024}. In what follows, we will report the entanglement dynamics of a set of double JC models where entanglement sudden death and revival will be shown under specific conditions.

For the present study, we assume both the cavities in the double JC model in the vacuum states $\ket{0_A}$ and $\ket{0_B}$.  We further assume that the two atoms are initially in the entangled states. We choose two such initial states of atoms~\cite{GDS+20} where one does not show entanglement sudden death while the other does, in the dynamics governed by the Hamiltonian~(\ref{eq:doubleJC,RWA}). 
The initial state of the total system in the former case is 

\begin{equation}
\ket{\psi_0} = (\sin(\alpha)\ket{l_A,e_B} + \cos(\alpha)\ket{e_A,l_B}) \otimes \ket{0_A,0_B},
\label{eq:no_sudden_death_state}
\end{equation}

while in the latter case, it is

\begin{equation}
\label{eq:ESD_int_state}
\ket{\tilde{\psi}_0} = (\sin(\alpha)\ket{l_A,l_B} + \cos(\alpha)\ket{e_A,e_B}) \otimes \ket{0_A,0_B}.
\end{equation}

We have summarized these cases in Appendix~\ref{app:analyticJC} with corresponding analytical expressions. These cases will be referred to as single excitation case and double excitation case respectively in the rest of the text.

\subsection{JC models in the presence of cavity disorders and noise}
\label{sec:disorder&noise}

Cavity disorders and noise are two practical aspects that may be incorporated into the studies of JC models. Noise models are frequently discussed, but cavity disorders are relatively less discussed. The origin of cavity disorder can be visualized, for instance, as a manifestation of the effect of the position of an atom inside a cavity leading to fluctuations in coupling strength $g$. Here, we focus on the glassy JC model or quenched disorder. In this case, the disorder remains effectively unchanged during the time for a single observation. The disorder equilibrates over a time scale much larger than the observation time being considered. The quantity of interest is obtained after quenched averaging numerous realizations of the disordered system.

\subsubsection {JC models with cavity disorders }

Disorders modeled as fluctuations in the cavity coupling strength ($g$) can be included in JC models. Thus, the interaction Hamiltonian (for double JC resonant case) in the presence of a cavity disorder becomes~\cite{GDS+20}

\begin{equation}
    \begin{split}
    \hat{\mathcal{H}} = & \frac{1}{2} \hbar \omega_0 \hat{\sigma}_3^A + \hbar \omega \hat{a}^\dagger \hat{a} + \hbar (1+\delta_A) G_A (\hat{\sigma}_+^A \hat{a} + \hat{\sigma}_-^A \hat{a}^{\dagger}) \\ & + \frac{1}{2} \hbar \omega_0 \hat{\sigma}_3^B + \hbar \omega \hat{b}^\dagger \hat{b} + \hbar (1+\delta_B) G_B (\hat{\sigma}_+^B \hat{b} + \hat{\sigma}_-^B \hat{b}^{\dagger}).
    \end{split}
    \label{eq:disorder-doubleJC}
\end{equation}

A simple comparison between Eqs.~(\ref{eq:doubleJC,RWA}) and~(\ref{eq:disorder-doubleJC}) would reveal that $\delta_{\{AbB\}}$ in Eq.~(\ref{eq:disorder-doubleJC}) quantifies the fluctuation in the coupling strength $G_{\{A,B\}}$. Here, $\delta_A$ and $\delta_B$ are the fluctuations in the cavity strengths for the atom-cavity system A and the atom-cavity system B respectively. Interestingly, there are different types of cavity disorders, which primarily differ from each other in the probability distribution from which $\delta$ is chosen. However, mainly two types of cavity disorders, namely \textit{Gaussian quenched disorder} and \textit{uniform quenched disorder}, have recently been studied~\cite{GDS+20}. These two types of cavity disorders can be briefly described as follows.

\begin{itemize}
    \item \textbf{Gaussian quenched disorder:} In this case, $\delta$ is chosen at random from a Gaussian distribution with zero mean and a standard deviation of $s$. The probability distribution function is given by   
    \begin{equation}
        P(\delta) = \frac{1}{s\sqrt{2\pi}} e^{-\frac{1}{2}\left(\frac{\delta}{s}\right)^2}.
     \end{equation}
    
    \item \textbf{Uniform quenched disorder:} In this case, $\delta$ is distributed uniformly between $\left[-\frac{s}{2},\frac{s}{2}\right]$. The probability distribution is given as 
    \begin{equation}
        P(\delta) = \begin{cases}
                        1/s, \, \mathrm{when} \, -\frac{s}{2}\leq \delta \leq \frac{s}{2}\\
                        0, \,  \mathrm{otherwise}.
                    \end{cases}
    \label{eq:uniform_dist}
    \end{equation}
\end{itemize}

Here, we study the impact of these two types of quenched disorders on the dynamics of the double JC model. To do so, we require quenched averaging of the quantities under consideration.
Specifically, the averaging is carried out after all other operations have been executed. For example, for quantifying the atom-atom entanglement (say in terms of concurrence $C$) in the double glassy JC system, the concurrence $C_\delta(t)$ is calculated for the atom-atom state $\psi_\delta(t)$ at a fixed time $t$. The quenched averaging of concurrence for the double JC case can be defined as

\begin{equation}
C(t) = \int_{-\infty}^{+\infty}\int_{-\infty}^{+\infty}C_{\delta_A, \delta_B}(t)P(\delta_A,\delta_B)(t)d\delta_Ad\delta_B.
\end{equation}

Numerically, $N$ instances of the disorder $\delta$ are Haar uniformly generated~\cite{ZK1994,TBW+2022}, and are labeled as $\delta^i$ and the quenched average value is calculated as

\begin{equation}
C(t) = \frac{1}{N^2} \sum_{i=1}^{N}\sum_{j=1}^{N}C_{\delta_A^i, \delta_B^j}(t).
\end{equation}

This approach allows us to investigate the dynamics of the process in the presence of probabilistic disorders/defects in the setup. Yet another non-ideal behaviour in the JC system can be its inevitable interaction with the ambient environments leading to dissipation and decoherence of the quantum features of the system, viz. entanglement. This can be studied using the master equation in open quantum system formalism.

\subsubsection{Open quantum system}\label{sec:open-syatem}

The system and environment/reservoir, initially uncorrelated, get correlated due to combined unitary evolution governed by a system-environment Hamiltonian. The system dynamics can be obtained by tracing out the environment as $\hat{\rho} = {\mathrm {Tr}}_{\mathrm{env}}[\hat{\rho}_{\mathrm{tot}}]$. In the Born-Markov approximations, the reduced dynamics of the system ($\rho$) can be described by the master equation~\cite{GZ10,BP02}

\begin{equation}
    \frac{d\hat{\rho}(t)}{dt} = -\frac{i}{\hbar}[\hat{\mathcal{H}}(t),\hat{\rho}(t)] + \sum_n \frac{\gamma_n}{2} [2\hat{A}_n\hat{\rho}(t)\hat{A}_n^\dagger - \hat{\rho}(t)\hat{A}_n^\dagger \hat{A}_n - \hat{A}_n^\dagger \hat{A}_n\hat{\rho}(t)],\label{eq:MEq}
\end{equation}

where $\hat{A}_n$ are the operators through which the system couples to the environment, and $\gamma_n$ are the corresponding rates. Markov approximation entails that the time scale of the decay of the correlations of the environment is much smaller than the time scale of the dynamics of the system ($\tau_{\mathrm{sys}} \gg \tau_{\mathrm{env}}$).

In the case of the double JC model (say in the model described by Eq.~(\ref{eq:doubleJC,RWA})), decoherence in the system could be induced due to a set of processes.
\begin{itemize}
\item Leakage of photons from the cavity into its surroundings with the average photon number $n_{\mathrm{th}}$: $\hat{A}_j=\hat{a}$ and $\gamma_j=\kappa(1+n_{\mathrm{th}})$. 
\item Photons leaking into the
cavity due to thermal excitation: $\hat{A}_j=\hat{a}^\dagger$ and $\gamma_j=\kappa n_{\mathrm{th}}$.
\item The polarization decay of the atom from the excited state to the ground state: $\hat{A}_j=\hat{\sigma}_-$ and $\gamma_j=\gamma$.
\item The dephasing of the energy states of the atom: $\hat{A}_j=\hat{\sigma}_z$ and $\gamma_j=\gamma_\phi$.
\end{itemize}

% We have already mentioned that the effects of noise on the entanglement in JC models is a promising avenue of study. Specifically, the possibility of noise in the system resulting in entanglement sudden deaths and revivals in the dynamics is of particular interest. Before we proceed to describe the results of such studies, it will be apt to briefly define a measure of entanglement and the meaning of entanglement sudden death. It should be noted that there is no unique measure of entanglement, but concurrence is a widely used~\cite{HW97,W98} measure of entanglement. The same is used in this paper to measure the correlation between the two atoms when studying the dynamics of the system evolution. 

\section{Results and discussion}
\label{sec:results}

Here, we investigate the evolution of entanglement characteristics between two atoms in different cavities that share the entanglement resource initially. We study this evolution under different variants of the double JC model and the effects of imperfections and/or open quantum systems described in the previous section. We have systematically performed simulations of the double JC system using QuTiP~\cite{qutip_1,qutip_2} to study entanglement dynamics. The study is performed using the master equation solver (available in QuTiP simulator) capable of solving the time evolution of a state using the master equation in the Lindblad form. More information about it can be found in Appendix~\ref{app:mesolve}. We focus on the emergence of entanglement sudden deaths and revivals as well as the effect of nonlinearity in influencing the time scale of the dynamics. To begin with, we discuss the case with the double JC model described through Eq. (\ref{eq:doubleJC,RWA}). 

The paper focuses on the strong coupling regime, where the damping rates of the optics and two-level system are much less than the coupling strength between them, i.e., $\kappa, \gamma << g$. The interactions studied in this paper also work in the regime where $g << \omega, \omega_0$. Optical experiments often fall in this regime, and this justifies the use of the rotating wave approximation (RWA) in simplifying the Hamiltonians of the interactions studied. The strong coupling regime is chosen for this study since strongly coupled systems are of interest in practical situations. This is so because it allows efficient energy transfer between the two coupled quantum systems on a time scale faster than the intrinsic noise of the system. Such strong coupling regimes have been achieved in cavity~\cite{TRG+92, RJM+01} and circuit quantum electrodynamics~\cite{WAD+04}, as well as in optomechanical~\cite{GKM+09, GMJ+19} and electromechanical~\cite{JDM+11} systems. 

\subsection{Double JC model}

\begin{figure}[b]
\centering
\subfloat[]{\begin{centering}
\includegraphics[width=0.49\textwidth]{ 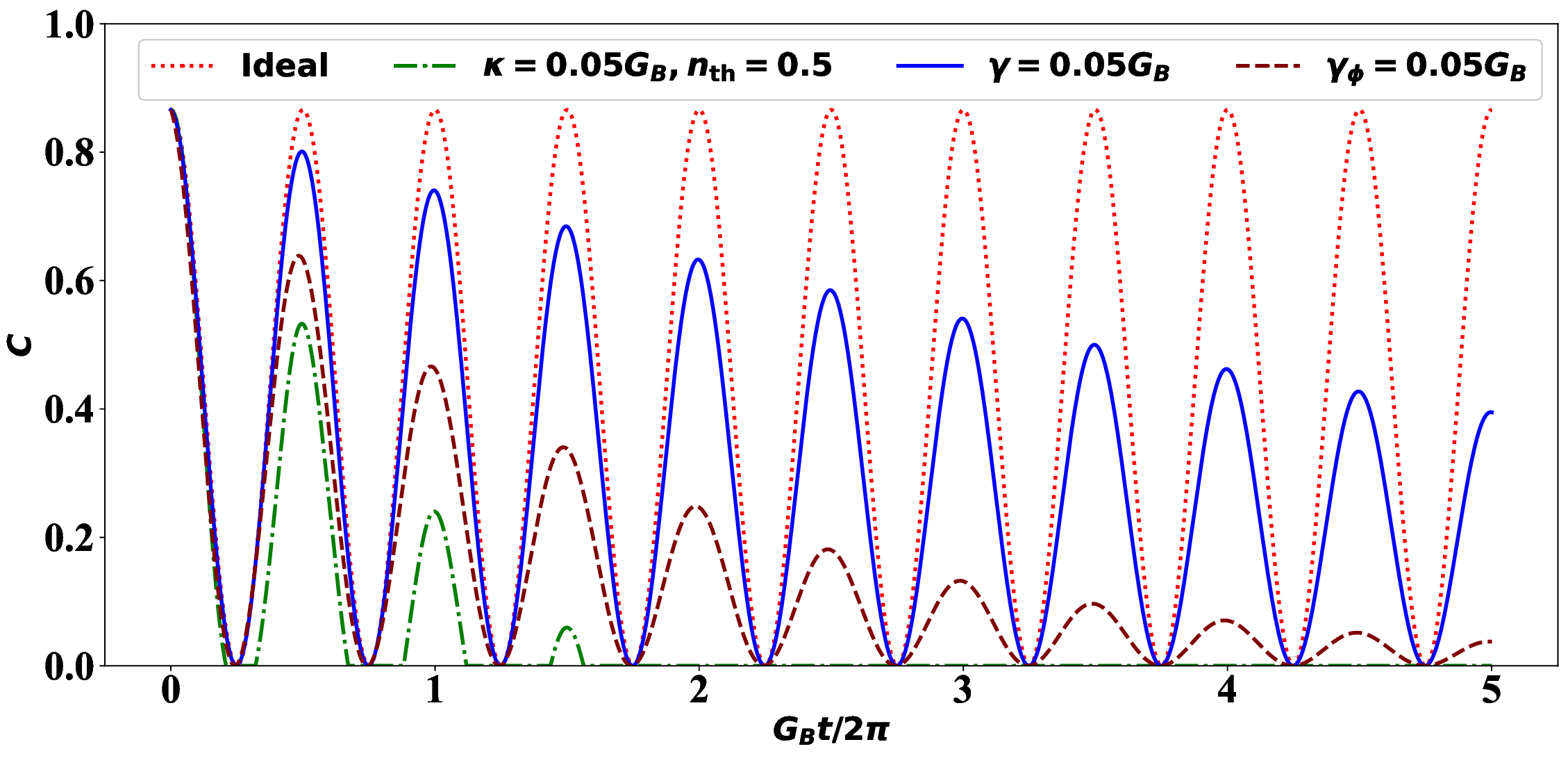}
\par\end{centering}
}
\subfloat[]{\begin{centering}
\includegraphics[width=0.49\textwidth]{ 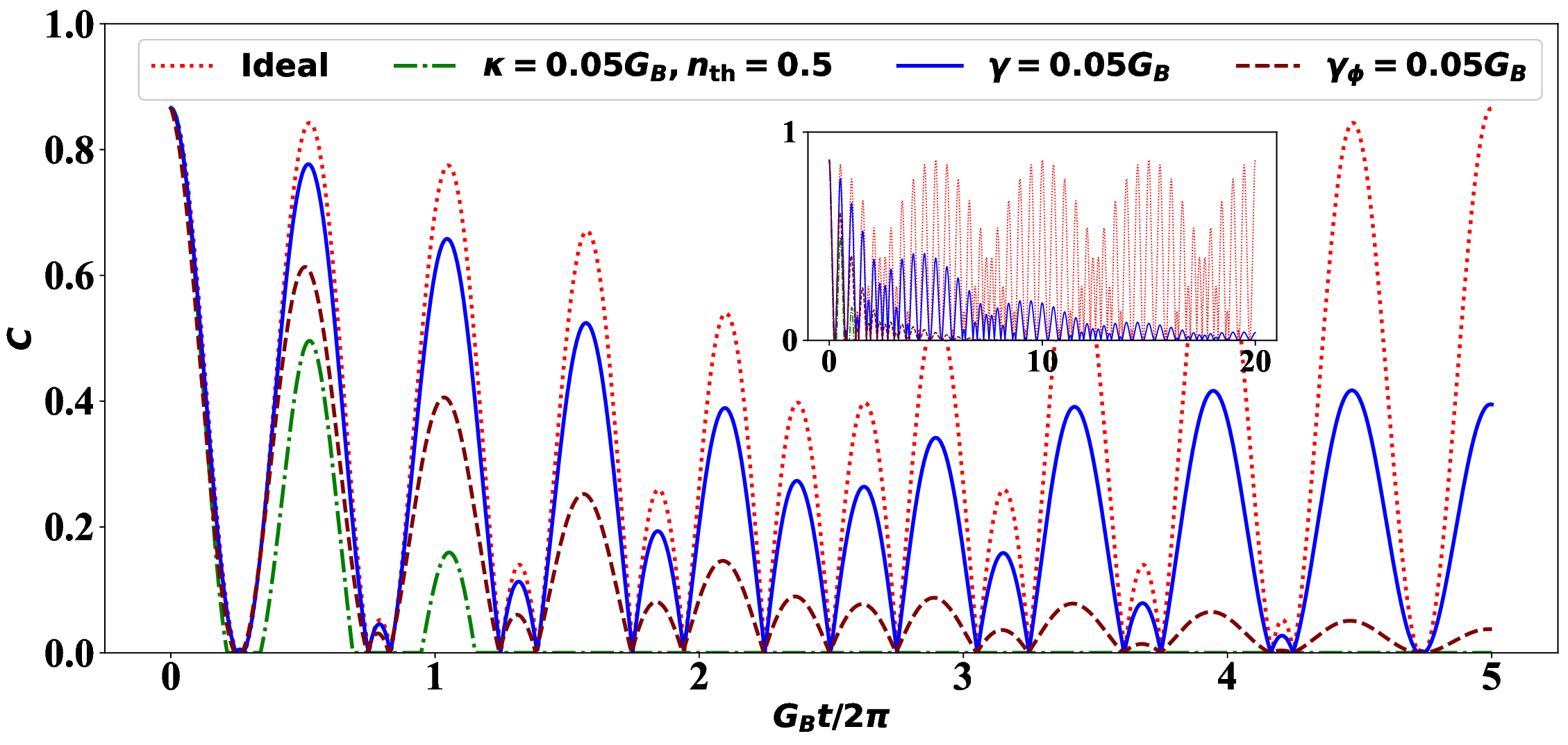}
\par\end{centering}
}
\\
\subfloat[]{\begin{centering}
\includegraphics[width=0.49\textwidth]{ 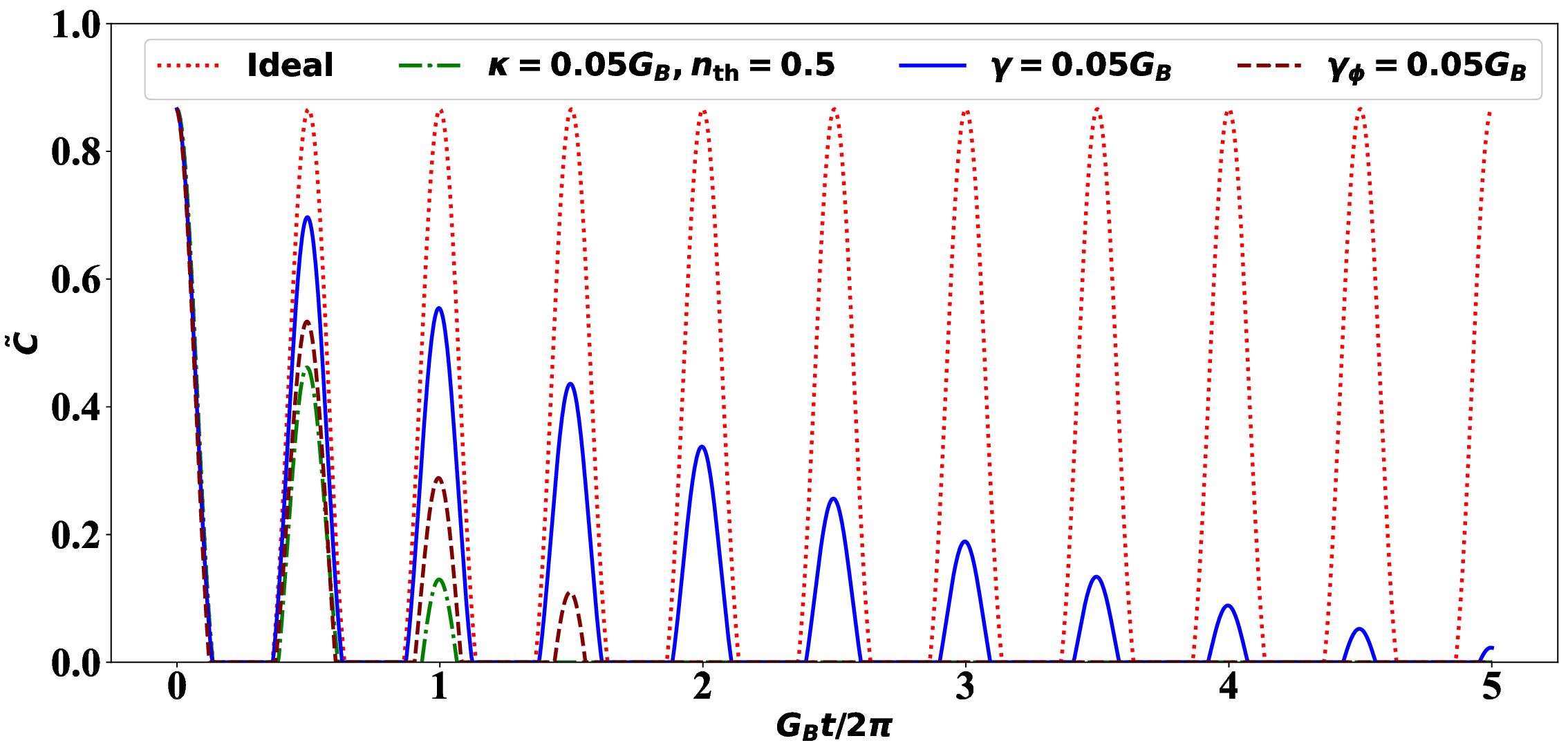}
\par\end{centering}
}
\subfloat[]{\begin{centering}
\includegraphics[width=0.49\textwidth]{ 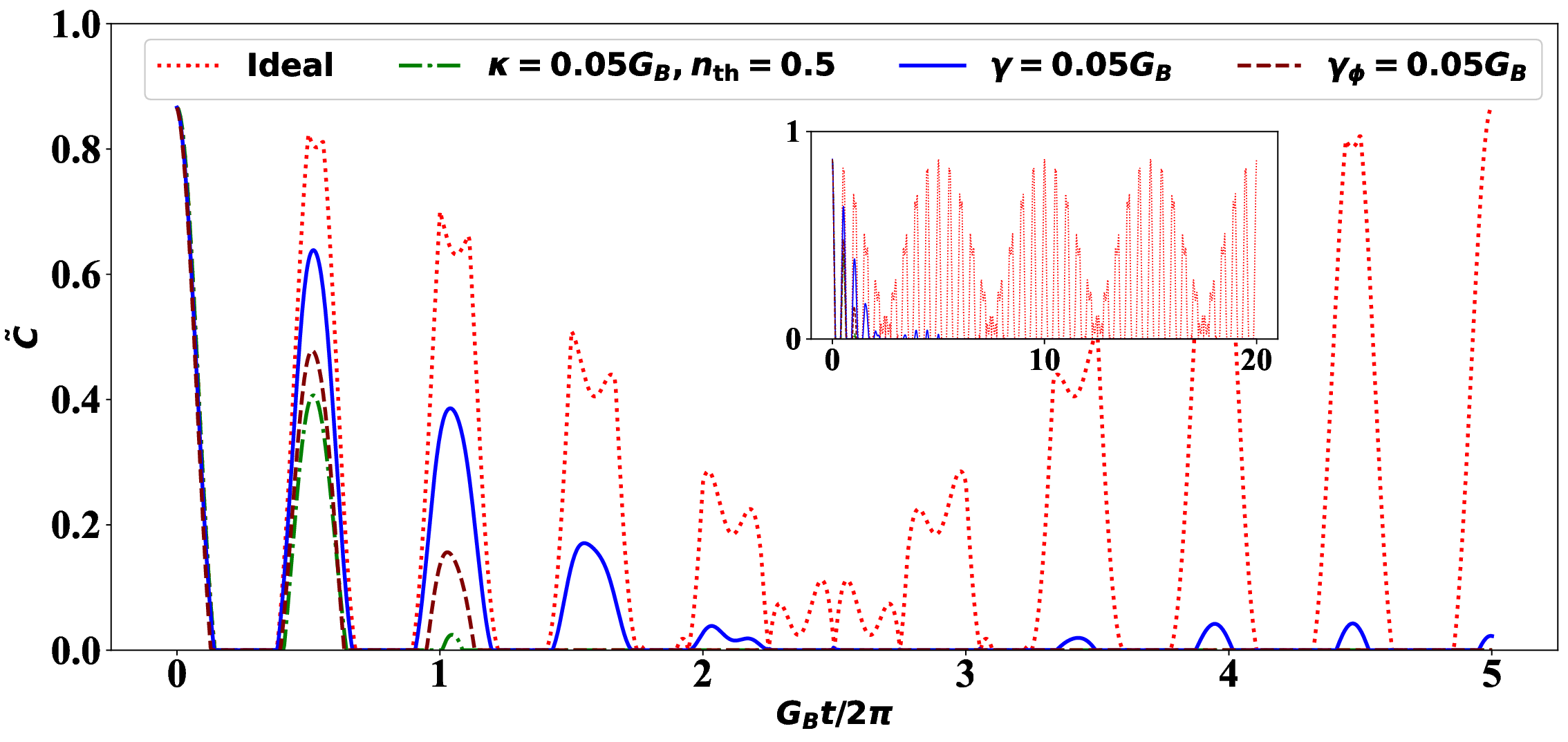}
\par\end{centering}
}
\caption{(Color online) Entanglement dynamics in double JC model in the single excitation case ($C$) is illustrated as (a) and (b) and the double excitation case ($\tilde{C}$) as (c) and (d). In (a) and (c),  symmetric cavities are considered, i.e., $G_A=G_B$, while asymmetry in the double JC model is considered as $G_A=0.9G_B$ in (b) and (d). The rest of the parameters are assumed to be the same for both the cavities, for example, $\kappa_A = \kappa_B = \kappa$, $\gamma_A = \gamma_B = \gamma$ and $\gamma_{\phi}^A = \gamma_{\phi}^B = \gamma_{\phi}$. Here and in what follows, the effect of noises is compared by studying them independently, such as the atomic decay without leakage and dephasing, dephasing without atomic decay and leakage, and cavity leakage without atomic decay and dephasing. Further, $\alpha = \frac{\pi}{6}$ and $\omega_0 = \omega = G_B$, i.e., resonant case is considered.}
\label{fig:dissipative_linear}
\end{figure}

We begin with the single excitation case where there are no entanglement sudden deaths and revivals present in the entanglement dynamics (under Hamiltonian~(\ref{eq:doubleJC,RWA})) in the absence of noise. In this case, we studied the dynamics of the system for different choices of coupling parameters with the initial state of the total system described by Eq.~(\ref{eq:no_sudden_death_state}). Here, we first discuss a symmetric case  ($G_A =G_B$) and an asymmetric case $G_A \neq G_B$. For illustration, we have specifically chosen $G_A = 0.9G_B$ for the asymmetric case. Evolution of the amount of entanglement (quantified in terms of entanglement measure concurrence) is shown in Figure~\ref{fig:dissipative_linear} (a) and (b) for symmetric and asymmetric cases, respectively, both in the absence and presence of various sources of dissipation discussed in Section~\ref{sec:open-syatem}. Here, we have used the same decay rates for both cavities, i.e., $\kappa_A = \kappa_B = \kappa$, $\gamma_A = \gamma_B = \gamma$ and $\gamma_{\phi}^A = \gamma_{\phi}^B = \gamma_{\phi}$. The regime considered here is the strong coupling regime (as $G_j > \kappa,\gamma,\gamma_{\phi}$)~\cite{FLR+2019,SB2009}.  The difference in the periodicity for the symmetric and the asymmetric case (in the ideal case) is apparent in the figure, with the asymmetric case being periodic at a time scale of about 10 times higher compared to the symmetric case. The periodic behavior of correlations between two resonant cavities has a similarity with the periodic behavior of atom-field entanglement in the single JC model~\cite{akella2022dynamics}. In the ideal asymmetric case, the dynamics reflect mirror symmetry along one-half of the period (cf. Figure~\ref{fig:dissipative_linear} (b)). Remnant of this behavior is visible for small decoherence rates. 

In the ideal case, we cannot observe entanglement sudden deaths and revivals, as the concurrence does not vanish with a discontinuous derivative. However, these deaths and revivals are present in the entanglement dynamics when the cavity interacts with a thermal reservoir, i.e., $n_{\mathrm{th}} > 0$. This can be attributed to the fact that the thermal bath injects non-coherent photons back to the dynamics with some probability, thus deteriorating the entanglement at lower thresholds. It is also noticeable (from the system Hamiltonian and dynamics) that the decay of entanglement with the dissipation of excitation through cavity (for vacuum bath) or atom (at the rate $\kappa$ or $\gamma$, respectively) has the same effect. This can be deduced by studying Eqs.~\ref{eq:singleJC,I,RWA} and \ref{eq:MEq}. Specifically, the interchange of the operators for atom dissipation ($\hat{\sigma}_{-}$) and the cavity dissipation ($\hat{a}$) leads to the same Hamiltonian. Thus, the Hamiltonian is symmetric in these operators for the linear case. Depletion of entanglement due to the dephasing of atomic states is more than that of dissipation of excitation at the same decay rates.  Due to the strong coupling regime, entanglement maxima at the consecutive peaks can be observed to decrease moderately with rescaled time. For a further small value of cavity decay rate and a thermal reservoir, the maxima of the entanglement after revival in the asymmetric case is observed to be higher than that before the entanglement sudden death. Decoherence finally eliminates all the quantum correlations between two cavities (shown in the inset of Figure~\ref{fig:dissipative_linear} (a) and (b)).    

We now consider the double excitation case, in which entanglement sudden deaths and revivals are observed (under Hamiltonian (\ref{eq:doubleJC,RWA})) even in the absence of noise. We can observe that at the instance of entanglement sudden death and revival, concurrence is continuous but not differentiable. An analytic expression showing the emergence of entanglement sudden death in a damped two qubit system is provided in the past~\cite{TJ2006}. The combined effects of amplitude and phase noise lead to sudden death in entanglement dynamics. The initial state of the atoms (\ref{eq:ESD_int_state})
reveals similar observations to that in the previous case  (shown explicitly in Figure \ref{fig:dissipative_linear} (c) and (d)). However, these similarities are more pronounced in the symmetric case than in the asymmetric case.  
Specifically, we observe that the addition of noise adversely affects the dynamics of the system, i.e., the maximum amount of entanglement after revival decreases. However, dephasing and dissipation were also observed to enhance the time between the entanglement sudden deaths and/or revivals. 
This may be visualized (except dissipation due to thermal reservoir) as a phenomenon in which noise suppresses smaller entanglement amplitudes and consequently drowns out some of the revivals. A specific example can be seen in the asymmetric case (cf. Figure \ref{fig:dissipative_linear} (d)) around one-half of the period in case of dissipation due to a vacuum reservoir.

\subsection{Multi-photon double JC model}

\begin{figure}[h]
     \centering
 \subfloat[]{\begin{centering}
\includegraphics[width=0.49\textwidth]{ 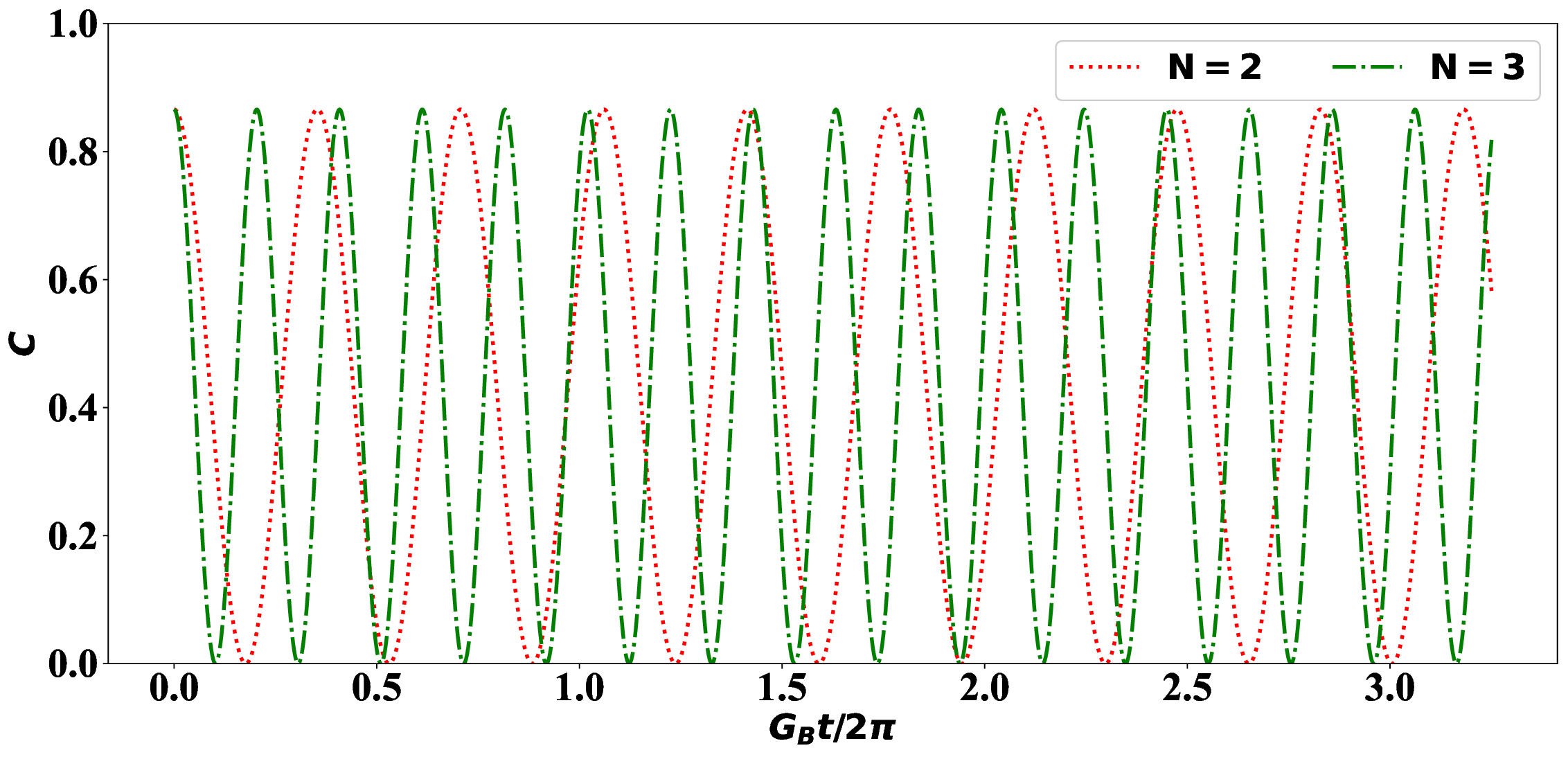}
\par\end{centering}
}    
 \subfloat[]{\begin{centering}
\includegraphics[width=0.49\textwidth]{ 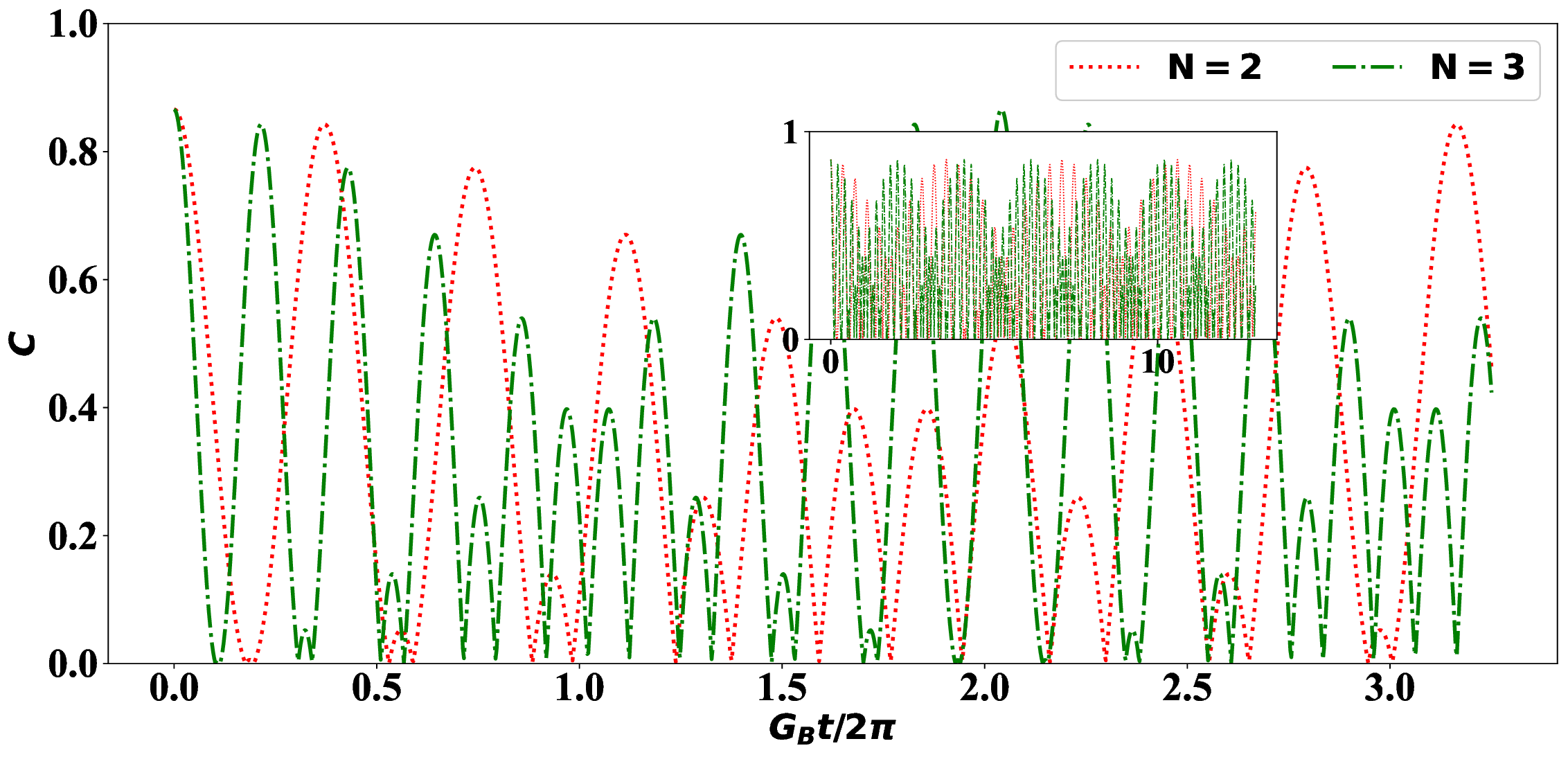}
\par\end{centering}
}    
\\
 \subfloat[]{\begin{centering}
\includegraphics[width=0.49\textwidth]{ 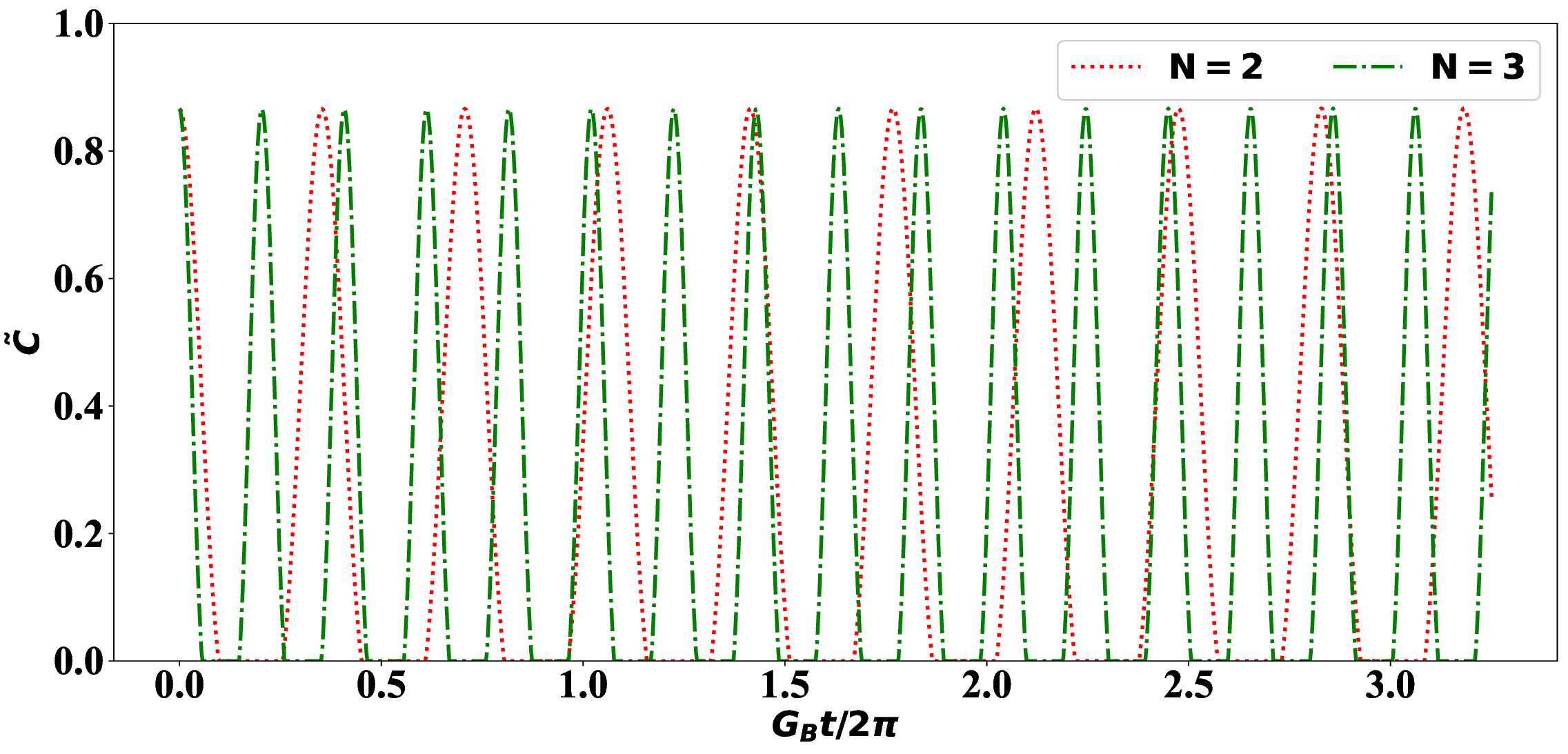}
\par\end{centering}
}    
 \subfloat[]{\begin{centering}
\includegraphics[width=0.49\textwidth]{ 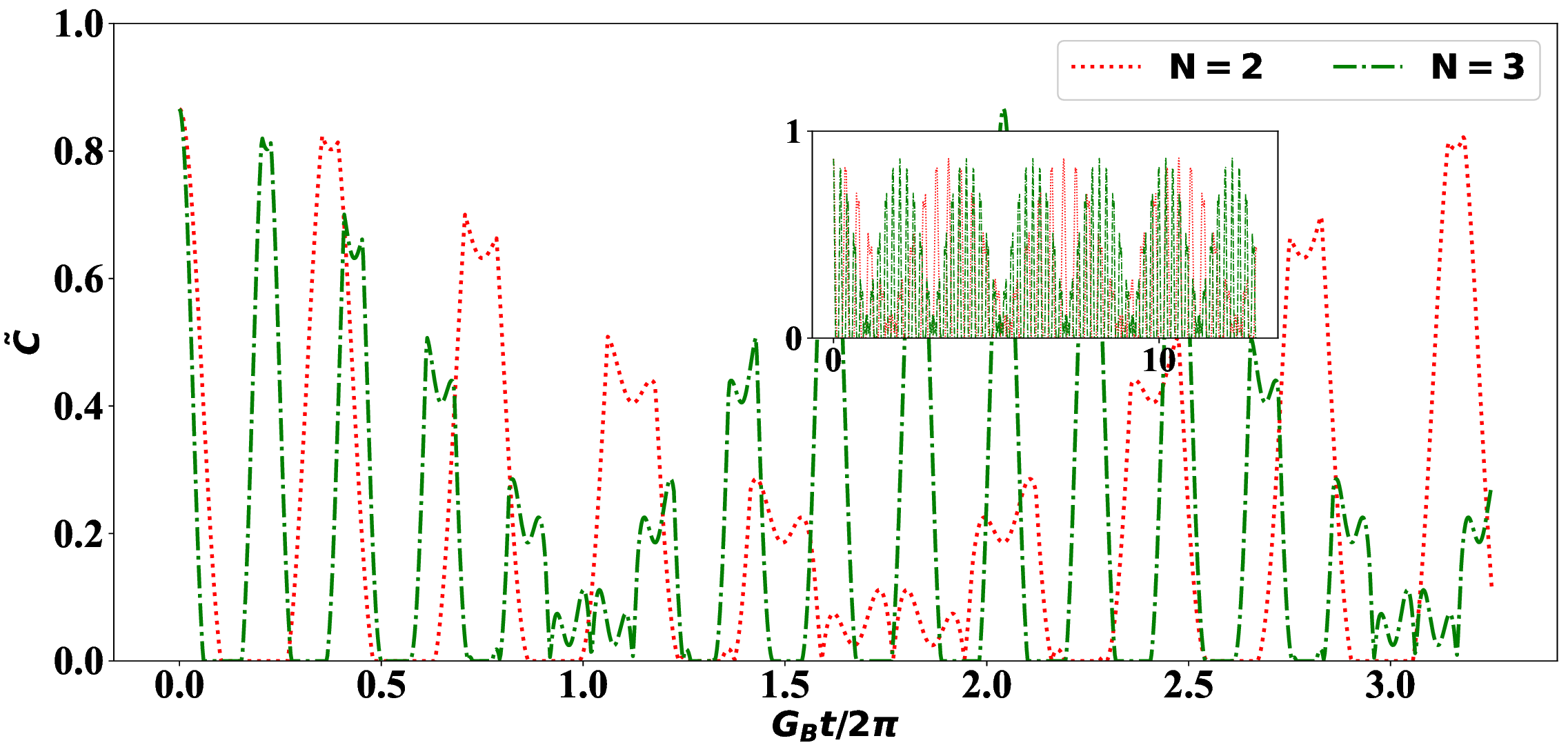}
\par\end{centering}
}   
\\
 \subfloat[]{\begin{centering}
\includegraphics[width=0.49\textwidth]{ 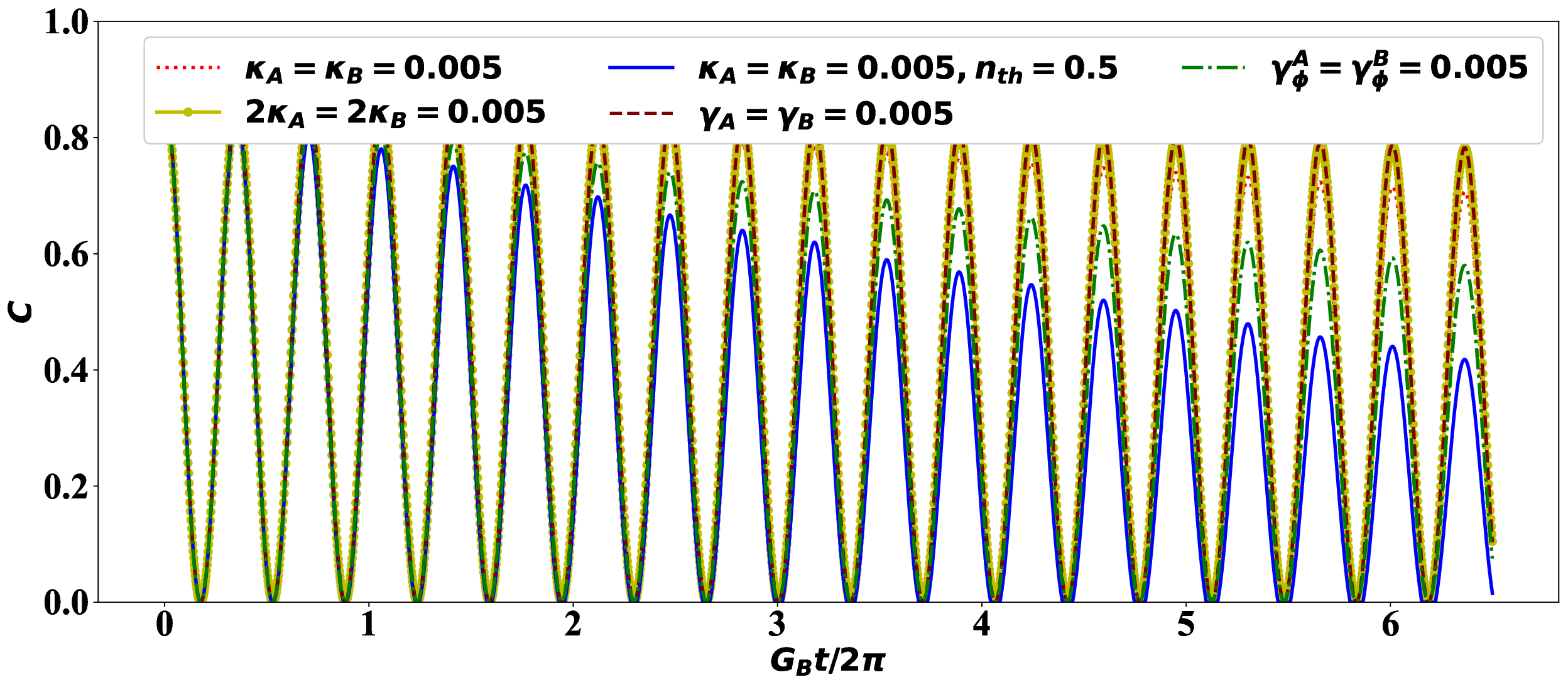}
\par\end{centering}
}    
 \subfloat[]{\begin{centering}
\includegraphics[width=0.49\textwidth]{ 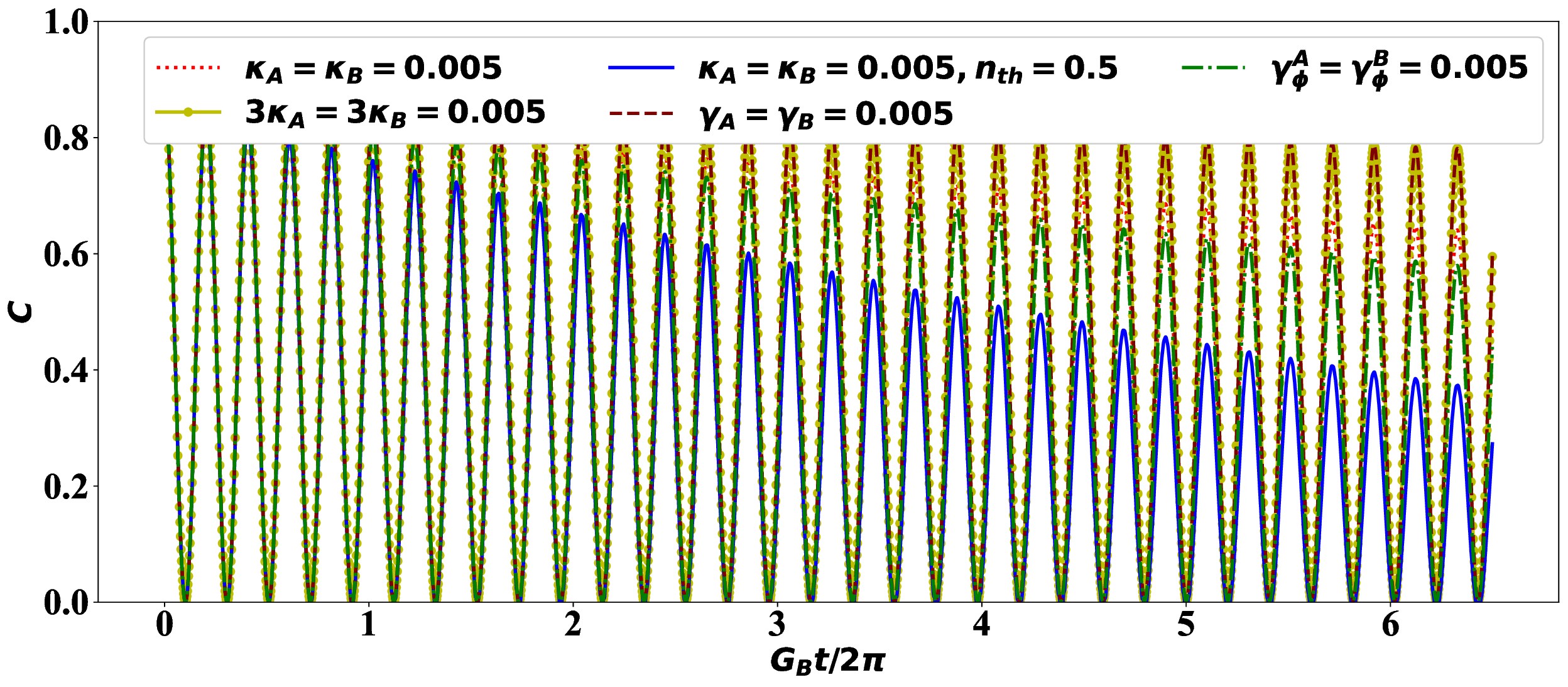}
\par\end{centering}
}           \caption{Entanglement dynamics in double JC model in the single excitation case ($C$) (in (a), (b), (e), and (f)) and the double excitation case ($\tilde{C}$) (in (c) and (d)). In (a), (c), (e), and (f),  symmetric cavities are considered, i.e., $G_A=G_B$, while asymmetry in the double JC model is considered as $G_A=0.9G_B$ in (b) and (d). Also, $N=2$ in (e) and $N=3$ in (f). {Dissipation rates are in units of $G_B$.} The rest of the parameters are assumed the same as in Figure~\ref{fig:dissipative_linear}.}

\label{fig:non_linear}
\end{figure}

We further study the entanglement dynamics of the multi-photon double JC model described by Eq.~(\ref{eq:doubleJC,non-lin,RWA}). The study of these nonlinear interactions in the strong coupling regimes is motivated by the Hamiltonian for photon-photon coupling presented in Refs.~\cite{SDE+18, SMA18}.
We begin with the ideal cases for which the results are shown in Figure \ref{fig:non_linear} (a)-(d). 
It is observed that higher-order of the multi-photon pump processes leads to faster system dynamics (and thus entanglement dynamics) compared to the corresponding lower-order case. For example, the period of the complete revival of the entanglement dynamics for the multi-photon double JC model for $N=3$ is smaller than that for $N=2$. This behavior is observed in both symmetric and asymmetric cavities. 
No other significant change is observed in the entanglement dynamics. 

When we introduce dissipation (noise) to our model, we start seeing more interesting dynamics as illustrated in Figure  \ref{fig:non_linear} (e) and (f). We restrict ourselves to the specific case where the initial state is given by Eq. (\ref{eq:no_sudden_death_state}) (i.e., the single excitation case) to focus on the effects of noise, disorder, and nonlinearity on introducing entanglement deaths and revivals. It is observed that the emergence of entanglement sudden deaths and revivals when setting $n_{th} > 0$ is preserved for the nonlinear cases. The onset of the sudden deaths and revivals is delayed with increasing nonlinearity (observed around scaled time $G_Bt/2\pi$ of $0.88$ in Figure ~\ref{fig:non_linear} (e),  compared to around $2.75$ in Figure \ref{fig:non_linear} (f)). Thus, it illustrates that a nonlinear interaction term makes the system robust against entanglement deaths and revivals in the presence of a thermal bath. Unlike the linear case in Figure \ref{fig:dissipative_linear}, where the decay of entanglement with the dissipation of excitation through the cavity (for vacuum bath) or atom (at the rate $\kappa$ or $\gamma$, respectively) has the same effect, this is not the case here, as is evident in Figures \ref{fig:non_linear} (e)-(f). Hence, the symmetry in the correlation dynamics due to $\kappa$ or $\gamma$ is broken. This can be understood by remembering that with the introduction of nonlinearity, the energy of the photons emitted due to the deexcitation of the atom is equivalent to the energy of a single photon in the cavity times the number of photons $N$ generated. The symmetry in the decay of correlation due to $\gamma$ and $\kappa$ is restored by scaling the $\kappa$ by $1/N$. The symmetry is also restored when scaling  $\gamma$ by $N$.

\subsection{Multi-photon double JC model in the presence of a nonlinearly interacting driving field}

\begin{figure}[h]
 \centering
 \subfloat[]{\begin{centering}
\includegraphics[width=0.49\textwidth]{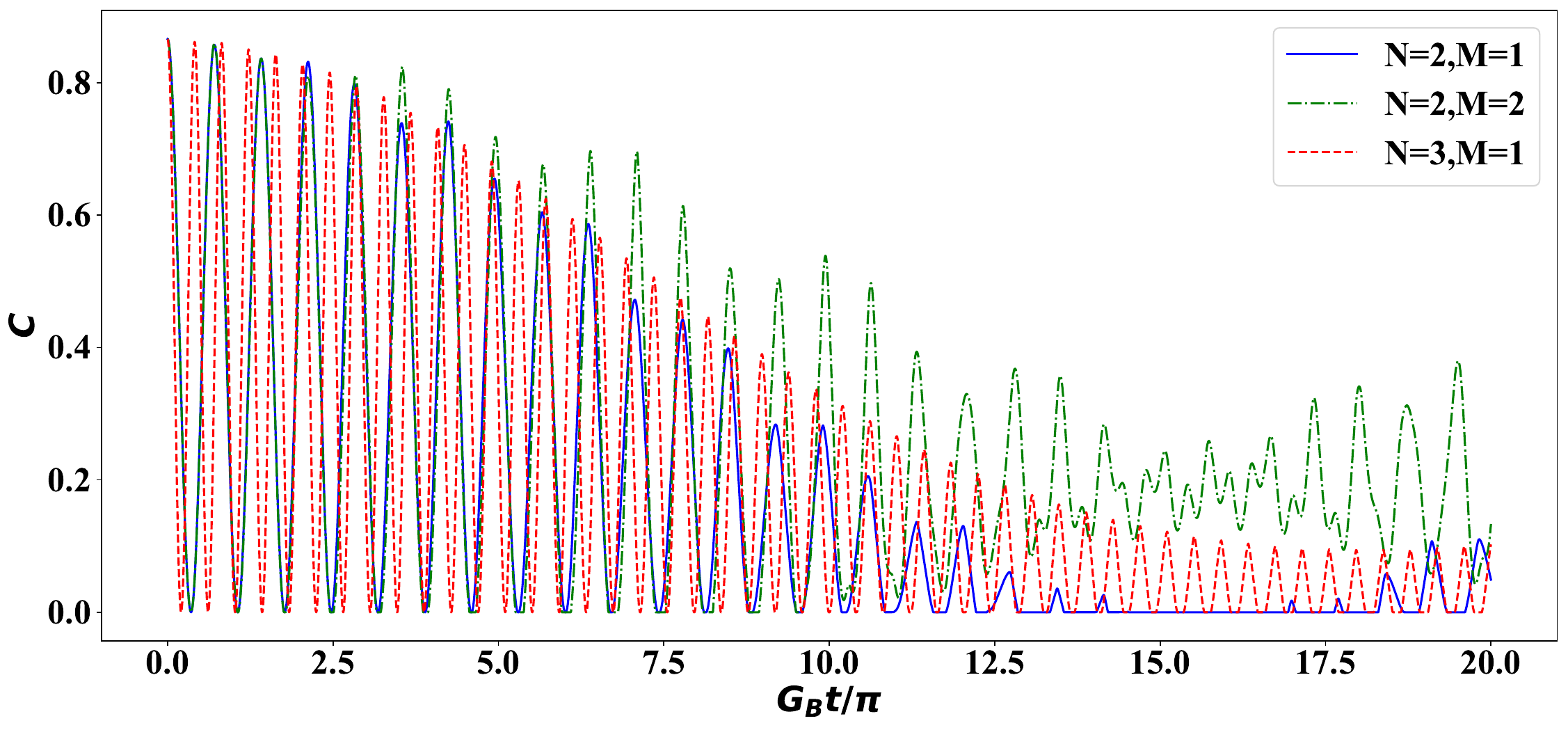}
\par\end{centering}
}   
 \centering
 \subfloat[]{\begin{centering}
\includegraphics[width=0.49\textwidth]{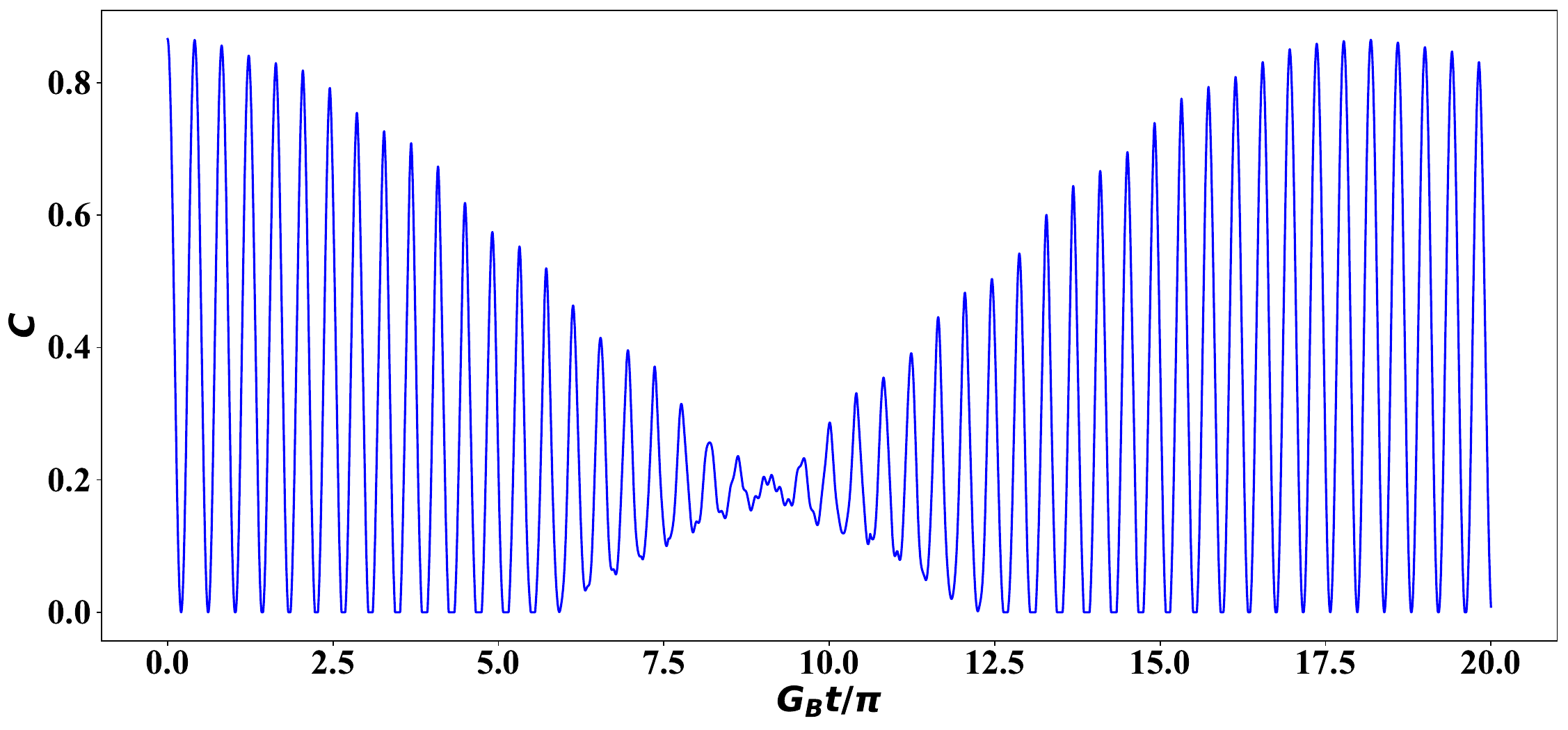}
\par\end{centering}
}  
     \caption{The evolution of concurrence in the nonlinear clean double JC case (without entanglement sudden death)  with the addition of a pump. Here, $\alpha = \frac{\pi}{6}$, $G_A = G_B$, and $\epsilon(t) = 0.01$ in Eqs.~\ref{eq:doubleJC,non_lin,pump}. In (a) {$(N,M) \in \{(2,1),(2,2),(3,1)\}$ and $N=3,M=3$} in (b). It is observed that the entanglement oscillates with time in the case of $M<N$. Entanglement sudden deaths and revivals are also observed even in the absence of dissipation. All the plots are generated at resonance.}
     \label{fig:non_linear_with_pump}
\end{figure}

\begin{figure}[h]
 \centering
 \subfloat[]{\begin{centering}
\includegraphics[width=0.49\textwidth]{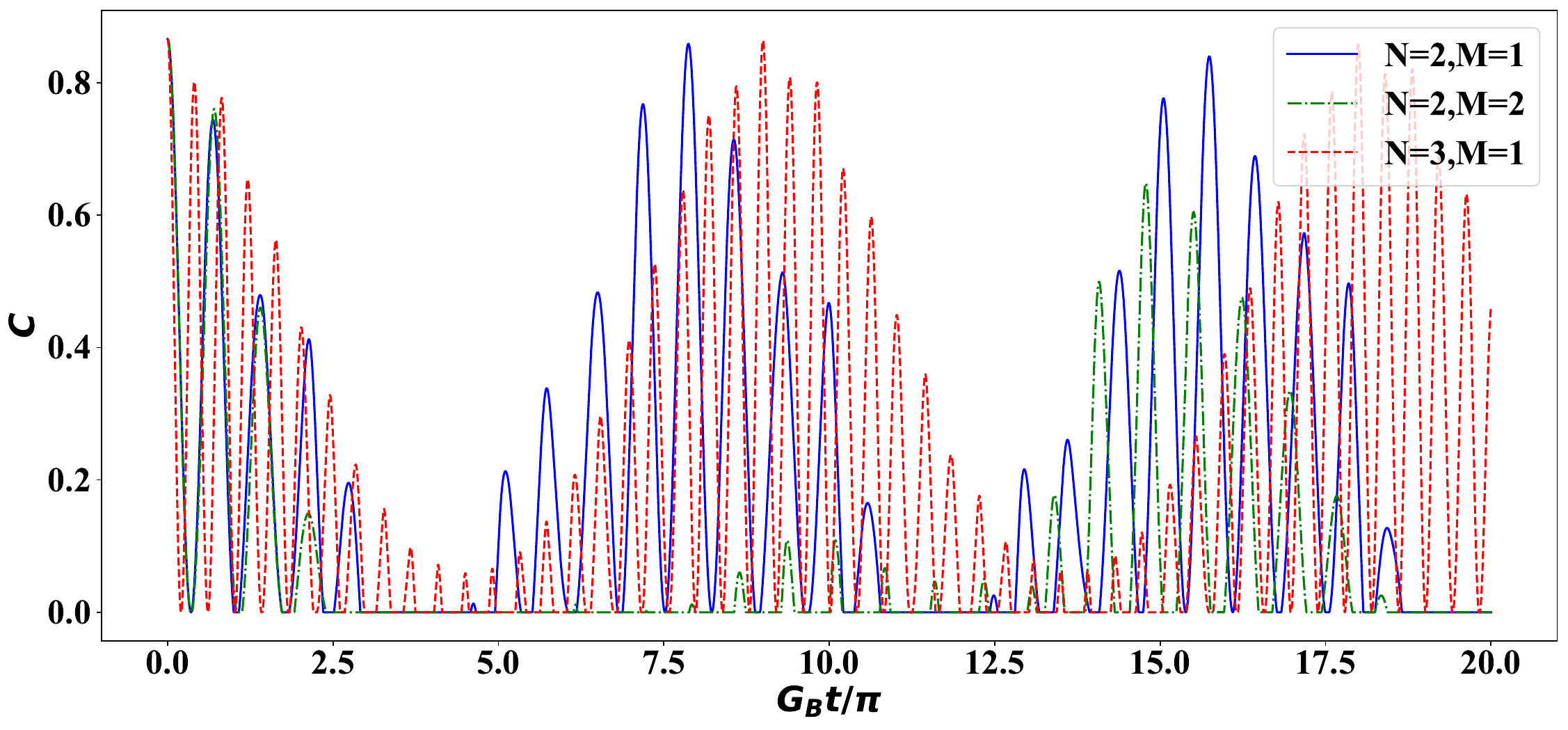}
        \par\end{centering}
}  
 \centering
 \subfloat[]{\begin{centering}
\includegraphics[width=0.49\textwidth]{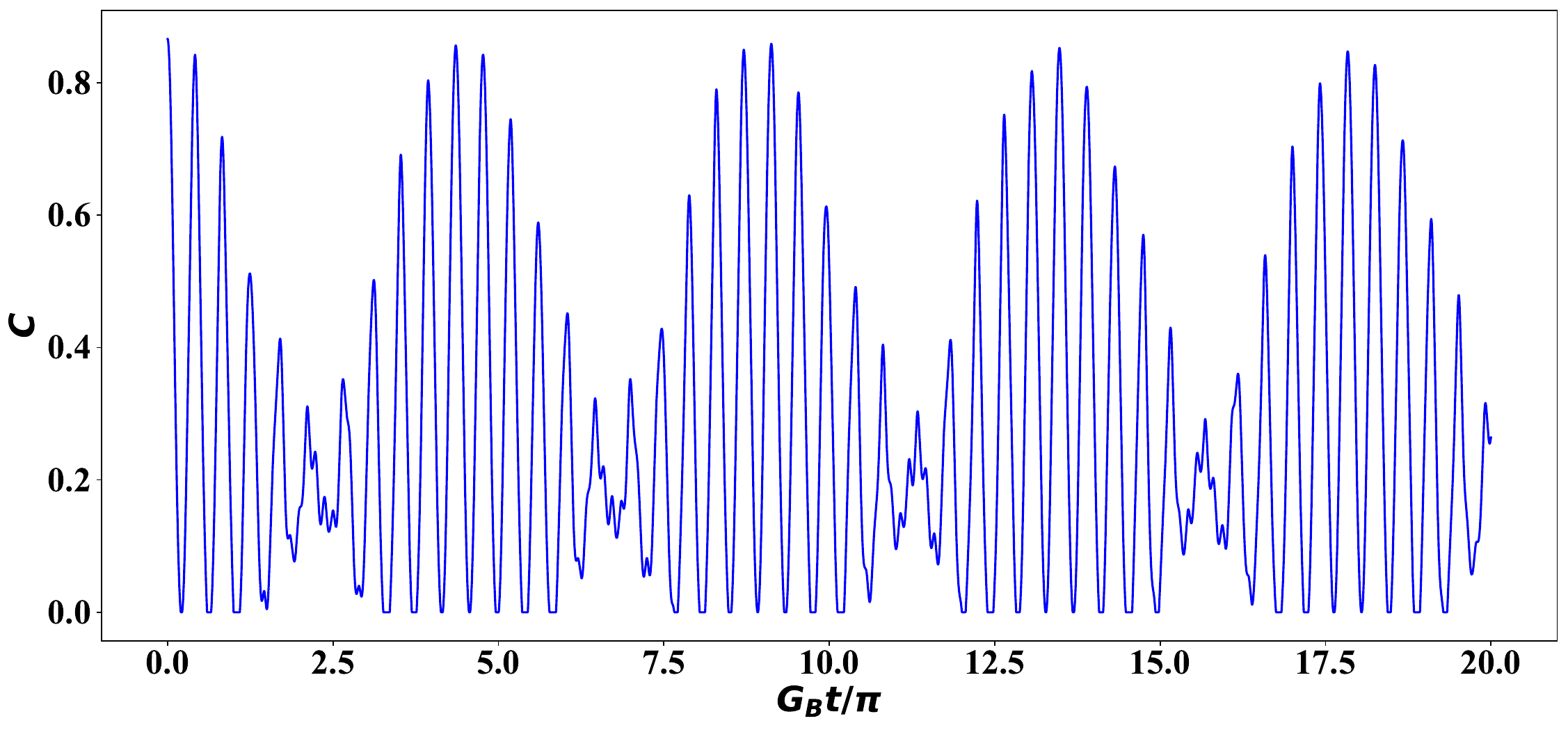}
        \par\end{centering}
}  
    \caption{Concurrence as in Figure \ref{fig:non_linear_with_pump} except the driving pulse is stronger at $\epsilon(t) = 0.04$. In (a) {$(N,M) \in \{(2,1),(2,2),(3,1)\}$ and $N=3,M=3$} in (b). It should be noted that the symmetry between the {$N=2,M=2$ and $N=3,M=3$} cases is broken when going from $\epsilon = 0.01$ in Figure \ref{fig:non_linear_with_pump} to $\epsilon = 0.04$.}
    \label{fig:non_linear_with_pump_2}
\end{figure}

In Figures ~\ref{fig:non_linear_with_pump} and \ref{fig:non_linear_with_pump_2}, we present the dynamics of the systems in the presence of a nonlinear pump. The Hamiltonian for this case under resonance is given by Eqs. \ref{eq:doubleJC,non_lin,pump}. Recent works by Hai-Ji Li et al.~\cite{HLS+2024} and Fen Zou et al.~\cite{zou2020multiphoton} have explored the effect of the combined effect of nonlinearity with the pump to study phonon blockade phenomena. One interesting aspect that can be noted is that adding a nonlinear pump gives rise to entanglement sudden deaths and revivals in the absence of noise, even for cases where such characteristics were not observed in the linear cases. For example, in Figure ~\ref{fig:non_linear_with_pump}, the characteristics of sudden death and revivals can be seen even in the absence of dissipation.  For a fixed strength of the driving field, increasing the nonlinearity alters the amount and periodicity of the concurrence. Higher orders of nonlinearity lead to periodic behavior in larger timescales. Another interesting observation to note here is that in Figure \ref{fig:non_linear_with_pump}, the cases for $N=M$ ($N=2, M=2$ and $N=3, M=3$) have a symmetry in the dynamics of the system. However when the strength of the driving field is increased, (cf. Figure \ref{fig:non_linear_with_pump_2}), this symmetry disappears. 

\subsection{Modelling Disorder}

\begin{figure}[h]
     \centering
 \subfloat[]{\begin{centering}
\includegraphics[width=0.49\textwidth]{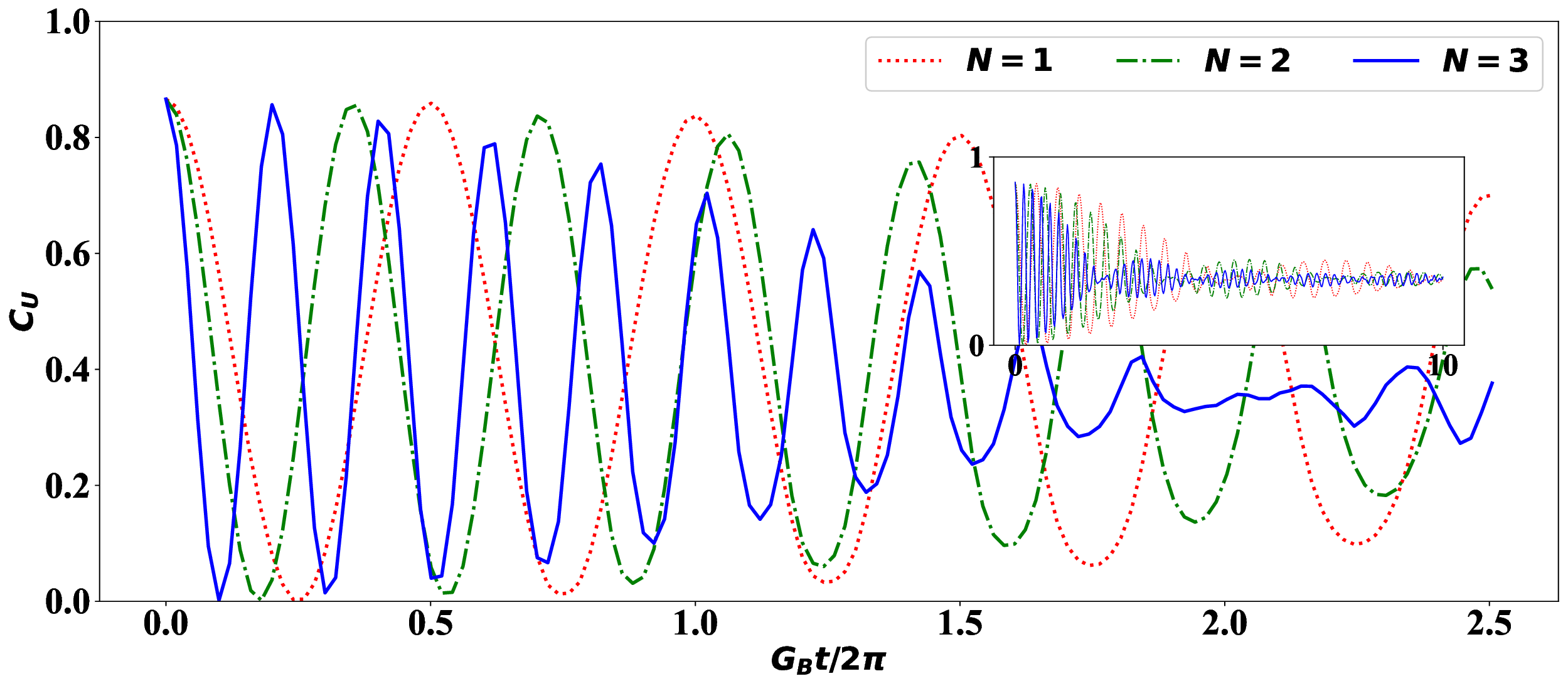}
          \par\end{centering}
} 
     \centering
 \subfloat[]{\begin{centering}
\includegraphics[width=0.49\textwidth]{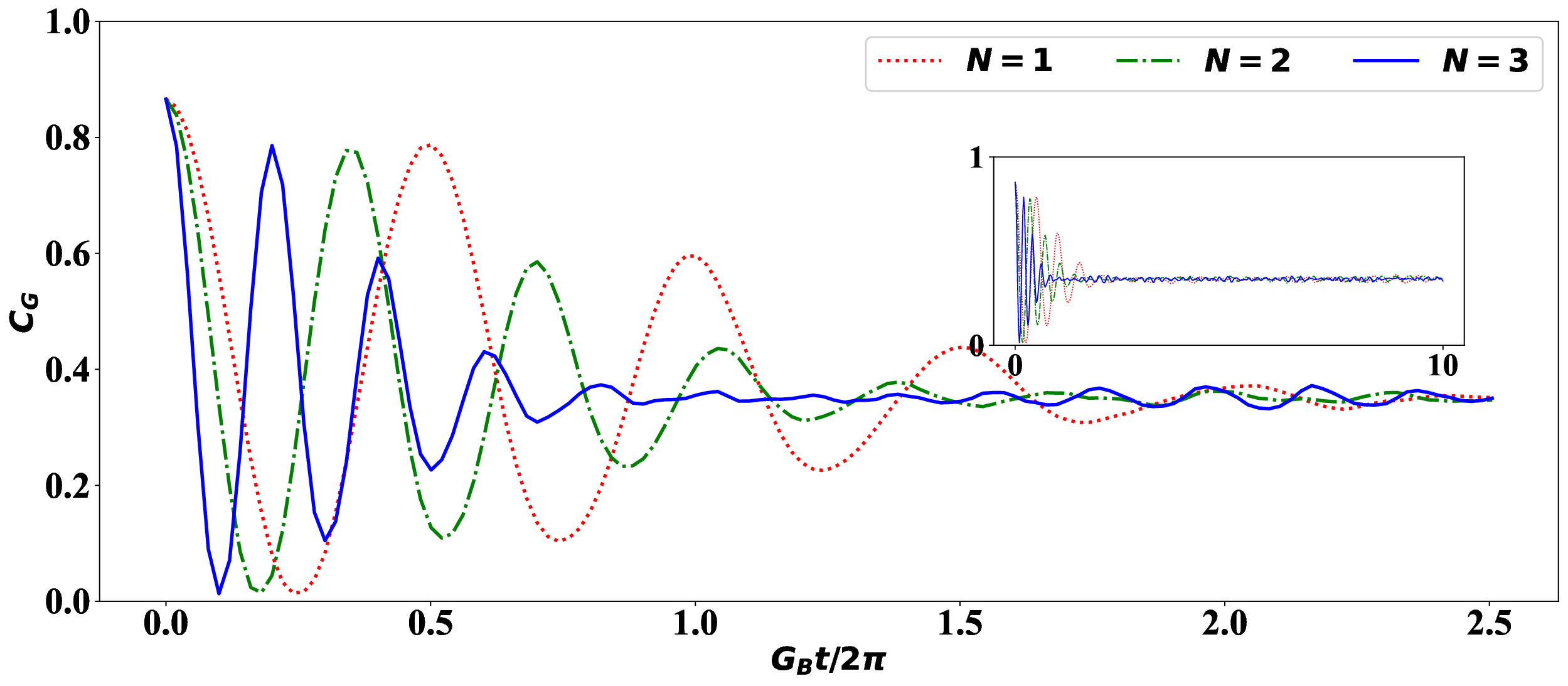}
          \par\end{centering}
} 
     \caption{The clean double JC model with  (a) uniform and (b) Gaussian disorders introduced for different values of nonlinearity ($N$ in the figure legend) with $G_A = G_B$. Also, $g_A=(1+\delta_A)G_A$ and $g_B=(1+\delta_B)G_B$ are selected independently at random for 1000 iterations according to the respective distribution of $\delta_A$ and $\delta_B$. It is seen that the dynamics of the system are faster for larger values of nonlinearity, which is consistent with earlier observations.}
     \label{fig:non_linear_disordered}
\end{figure}

\begin{figure}[h]
     \centering
 \subfloat[]{\begin{centering}
\includegraphics[width=0.49\textwidth]{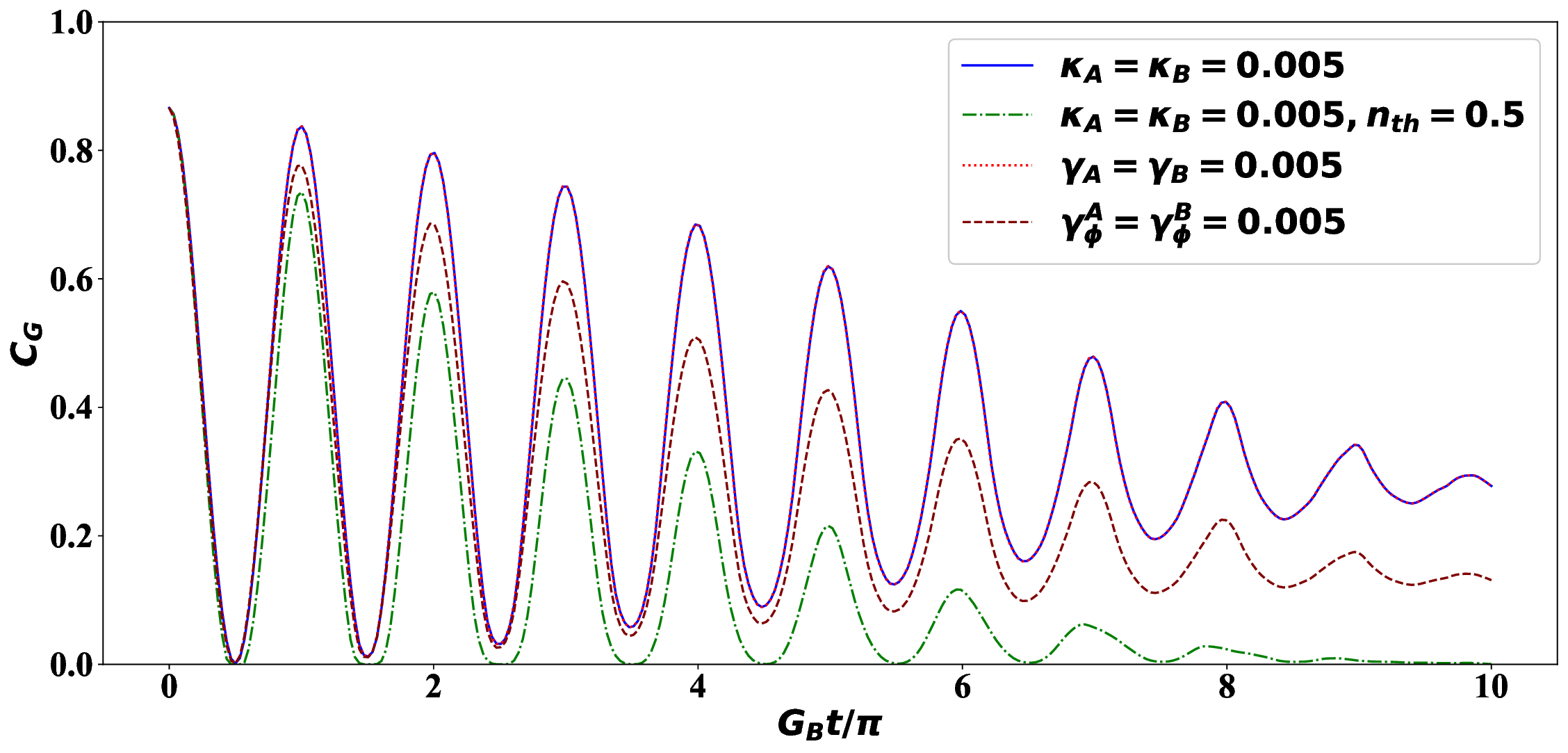}
          \par\end{centering}
} 
     \centering
 \subfloat[]{\begin{centering}
\includegraphics[width=0.49\textwidth]{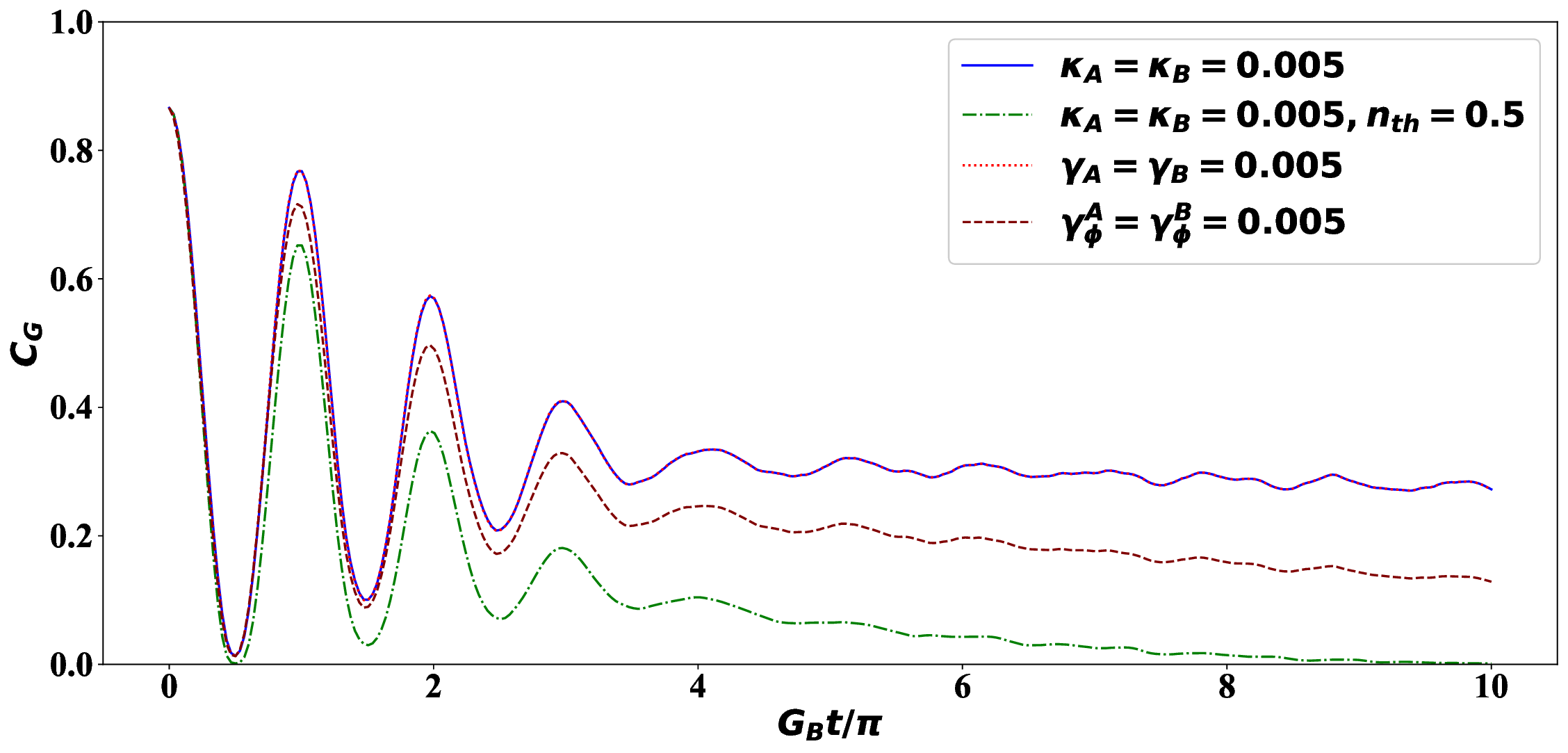}
          \par\end{centering}
} 
     \caption{The linear double JC model with (a) uniform and (b) Gaussian disorders introduced along with various dissipation due to noise with $G_A = G_B$. Also, $g_A=(1+\delta_A)G_A$ and $g_B=(1+\delta_B)G_B$. For (a) $\delta_A$ and $\delta_B$ are selected independently at random for 1000 iterations according to the uniform distribution given in Eq. (\ref{eq:uniform_dist}) for $s=0.1$. For (b) $\delta_A$ and $\delta_B$ are selected independently at random for 1000 iterations according to the normal distribution, with zero mean and 0.1 standard deviation. {Dissipation rates are in units of $G_B$.}}
     \label{fig:dissipative_disordered}
\end{figure}

We introduce different disorders to the clean double JC model without dissipation. We observe a trend similar to the one seen in Figure \ref{fig:non_linear} (a)-(d). 
Increase in the nonlinearity speeds up the dynamics of the system and the entanglement washes out faster as can be seen in Figure \ref{fig:non_linear_disordered}. Introducing dissipation (noise) along with the disorder also results in dynamics similar to the case without dissipation, but the characteristic amplitudes are found to decay as shown in Figure \ref{fig:dissipative_disordered}.

The main characteristic effect of introducing disorder on the dynamics is that it washes out the entanglement dynamics to its average value from the clean case. A similar observation on the dynamics of coherence in the presence of glassy disorders in the system is reported in~\cite{JFN+2023}. This can be inferred as the average value of the entanglement measure is obtained by averaging the dynamics of multiple systems with coupling strengths distributed according to a defined distribution. The different coupling strengths give rise to different timescales for the Rabi oscillations for the multiple systems. Thus, averaging over them leads to an average value for the entanglement over long timescales. The coherence timescales for the washing out of the dynamics are determined by the underlying distributions of the disorder. As seen from Figure \ref{fig:non_linear_disordered}, a Gaussian quenched disorder leads to shorter coherence time scales than a uniform distribution for the same variance in the distributions. This might be understood by the fact that a uniform distribution has a definite cutoff in the values of coupling strengths that are possible, a Gaussian distribution has a tail that extends to infinity.

\section{Conclusions}\label{sec:conclusions}
In this work, we have studied the entanglement dynamics of a set of double JC models, which includes models where the two atoms share an initial entanglement resource in the ideal case and under the presence of dissipation, nonlinear interaction with the cavity field, nonlinear driving field, and cavity disorders. These independent processes impact the dynamics of entanglement when they act simultaneously. For example, the effect of disorder and dissipation is studied in the case when multiphoton interactions are present between the cavity and the atom. The effects of competing processes are studied when both dissipation and disorder are present in the system, and the effect on the dynamics of introducing a nonlinear interaction term along with the noise is also studied. The evolution is calculated numerically using the QuTiP~\cite{qutip_1,qutip_2} framework. 

The effect of noise on the dynamics of the system is to deteriorate entanglement, while that of the disorder is to wash them out. On the other hand, the introduction of nonlinearity is found to cause the dynamics of the system to speed up. A combination of these effects gives rise to interesting phenomena, like entanglement sudden deaths, and revivals, even in cases where that did not exist before.

We found that the introduction of thermal noise leads to the appearance of sudden deaths and revivals in the entanglement dynamics even when they were absent in the noiseless case. Furthermore, the action of decay noise in the cavity field is similar to the atomic noise, i.e., they are interchangeable. In the cases where sudden death and revivals are present in the clean case, the time between them is increased with the introduction of environment-induced decoherence. The introduction of a nonlinear interaction term in the Hamiltonian delays the onset of entanglement sudden deaths in the dynamics. Introducing a nonlinear driving field also introduces entanglement sudden death and revival dynamics even without thermal photons. Nonlinearity helps in combating environment-induced decoherence. However, cavity disorder impacts the cases with higher-order nonlinearities by declining the corresponding entanglement faster. Therefore, the advantage of nonlinear atom-cavity coupling is suppressed due to cavity disorders.

As the paper reports the entanglement dynamics of a large family of double JC models in various conditions, it is expected to be of use in the study of various quantum phenomena where JC models are known to have relevance. Further, the present study is restricted to Markovian noise. It is expected that future studies along similar lines will reveal rich dynamics of the double JC model in the presence of non-Markovian noise. This is because, in the presence of non-Markovian noise, the time evolution depends on the state at previous times, giving rise to the 'memory effect' and time correlations. One possible way to straightforwardly extend the numerical methods used here to study non-Markovian noise is to use the ensemble of Linbaldian trajectories approach, which captures the non-Markovian dynamics by taking a weighted average of Linbaldian trajectories at different time lags~\cite{KD2019}.

\section*{Acknowledgments}
AP acknowledges the support from the QUEST scheme of Interdisciplinary Cyber Physical Systems (ICPS) program of the Department of Science and Technology (DST), India (Grant No.:
DST/ICPS/QuST/Theme-1/2019/14 (Q80)). HR acknowledges the help of Anthony K.C. Tan and Evan Cryer Jenkins in refining the structure of the manuscript. 

\textbf{Disclosures.} The authors declare no conflicts of interest.

\bibliography{references}

%\newpage

\appendix

\section{Solution of the double JC in the Clean Case}
\label{app:analyticJC}
 
The Hamiltonian of the double JC model is defined in Eq. (\ref{eq:doubleJC,RWA}), which leads to the time evolution 
of the initial state (\ref{eq:no_sudden_death_state}) by 
time-dependent Schr\"odinger equation
\begin{equation}
    \ket{\psi_t} = e^{\frac{-i\mathcal{H}t}{\hbar}}[ (\sin(\alpha)\ket{l_A,e_B} + \cos(\alpha)\ket{e_A,l_B}) \otimes \ket{0_A,0_B}].
\end{equation}
This can be further simplified to 
\begin{equation}
    \begin{split}
        \ket{\psi_t} = & \cos(\alpha) \{\cos(G_A t)\ket{e_A,0_A} - i\sin(G_At)\ket{l_A,1_A}\}\otimes \ket{l_B,0_B} \\ & + \sin(\alpha)\ket{l_A,0_A}  \otimes \{\cos(G_B t)\ket{e_B,0_B} - i \sin(G_B t)\ket{l_B,1_B}\}.
    \end{split}
\end{equation}
We can quantify the amount of entanglement between two atoms after tracing out the cavity modes as concurrence (\ref{eq:conC}). The reduced atom-atom quantum state is
 \begin{equation}
    \rho_{AB}(t) = 
    \begin{bmatrix}
        0 & 0 & 0 & 0 \\
        0 & a & p & 0 \\
        0 & p^* & b & 0 \\
        0 & 0 & 0 & 1-a-b \\
    \end{bmatrix},
\end{equation}
with 
\begin{equation*}
    \begin{split}
        a & = \cos^2(\alpha)|\cos(G_At)|^2 \\
        b & = \sin^2(\alpha)|\cos(G_Bt)|^2 \\
        p & = \cos(\alpha)\sin(\alpha)\cos(G^*_At)\cos(G_Bt)
    \end{split}
\end{equation*}
which gives us atom-atom entanglement 
\begin{equation}
    C(t) = |\sin(2\alpha)\cos(G_At)\cos(G_Bt)|.\label{eq:C1}
\end{equation}

Similarly, the initial state (\ref{eq:ESD_int_state}) evolves to 
\begin{equation}
    \ket{\tilde{\psi}_t} = e^{\frac{-i\hat{\mathcal{H}}t}{\hbar}}[(\sin(\alpha)\ket{l_A,l_B} + \cos(\alpha)\ket{e_A,e_B}) \otimes \ket{0_A,0_B}],
\end{equation}
which can be simplified to 
\begin{equation}
    \begin{split}
        \ket{\tilde{\psi}_t} = & [\cos(\alpha) \{\cos(G_At)\ket{e_A,0_A} - i\sin(G_At)\ket{l_A,1_A}\} \\ & 
        \otimes 
        \{\cos(G_Bt)\ket{e_B,0_B} - i\sin(G_Bt)\ket{l_B,1_B}\} 
        + \sin(\alpha) \ket{l_A,0_A}  \otimes \ket{l_B,1_B}].
    \end{split}
\end{equation}
Entanglement between two atoms in the quantum state 
\begin{equation}
    \tilde{\rho}_{AB}(t) = 
    \begin{bmatrix}
        e & 0 & 0 & h^* \\
        0 & f & 0 & 0 \\
        0 & 0 & g & 0 \\
        h & 0 & 0 & 1-e-f-g \\
    \end{bmatrix},
\end{equation}
where 
\begin{equation*}
    \begin{split}
        e & = \cos^2(\alpha)|\cos(G_At)\cos(G_Bt)|^2 \\
        f & = \cos^2(\alpha)|\cos(G_At)\sin(G_Bt)|^2 \\
        g & = \cos^2(\alpha)|\sin(G_At)\cos(G_Bt)|^2 \\
        h & = \cos(\alpha)\sin(\alpha)\cos(G_At)\cos(G_Bt)
    \end{split}
\end{equation*}
after tracing out the cavity modes as quantified using concurrence (\ref{eq:conC})
\begin{equation}
        \tilde{C} =  \max\{0,|\sin(2\alpha)\cos(G_At)\cos(G_Bt)|  - \frac{1}{2}\cos^2(\alpha)|\sin(2G_At)\sin(2G_Bt)|\}.
\label{eq:C2}
\end{equation}
We can observe from Eqs. (\ref{eq:C1}) and (\ref{eq:C2}) that the former quantity remains positive for all the values of parameters; while in the latter case, it becomes zero whenever the $|\sin(2\alpha)\cos(G_At)\cos(G_Bt)|  < \frac{1}{2}\cos^2(\alpha)|\sin(2G_At)\sin(2G_Bt)|$. This gives the entanglement sudden death in the atom-atom entanglement evolution from the initial state (\ref{eq:ESD_int_state}). 
For more details of the ideal case and the role disorders on the dynamics of entanglement please refer to (\cite{YYE06,GDS+20} and references therein).

\section{Qutip Master Equation Solver Methodology}
\label{app:mesolve}
We utilize the QuTiP quantum toolbox~\cite{qutip_1,qutip_2} to simulate and solve the Lindblad Master equation given in Section~\ref{sec:open-syatem}. The evolution is calculated numerically using Linear Multistep Methods. The equations can be solved using either the Adams method~\cite{HNW93} or the Backward differentiation formulae (BDF)~\cite{CH52}. Adams methods are suitable for non-stiff differential equations, whereas BDF methods are suitable for stiff differential equations. We used the former method for integration,  which utilizes scipy with the 'zvode' integrator. It is a variable-coefficient ordinary differential equation solver, with fixed-leading-coefficient implementation. The options used for the integrator are given in Table~\ref{tab:integrator_params}. Interested readers can check the 'zvode' documentation ~\cite{ZVODE} and the source code for the QuTiP master equation solver ~\cite{QuTiP_mesolve}.

\addtolength{\tabcolsep}{+16pt}
\begin{table}[ht]
\caption{ZVODE Integrator Parameters}
\centering 
\begin{tabular}{l c c} 
\hline\hline 
\textbf{Parameter} & \textbf{Value} \\ [0.5ex] 
\hline 
Absolute Tolerance & 1e-8 \\
Relative Tolerance & 1e-6 \\
Solver Method      & 'adams' \\
Order of Solver    & 12 \\
Max. number of steps \\ to take for each interval & 1000 \\
Size of initial step & Determined Automatically \\
Minimum Step Size & Determined Automatically \\
Maximum Step Size & Determined Automatically \\
\\
[1ex]
\hline \hline
\end{tabular}
\label{tab:integrator_params}
\end{table}

\end{document}